\documentclass[final,5p,times,10pt,authoryear,twocolumn]{elsarticle}

\newcommand\aap{A\&A}                
\newcommand\aj{AJ}                   
\newcommand\apj{ApJ}                 
\newcommand\apjl{ApJ}                
\newcommand\apjs{ApJS}               
\newcommand\araa{ARA\&A}             
\newcommand\mnras{MNRAS}             
\newcommand\nat{Nature}              
\newcommand\prd{Phys. Rev.~D}        
\newcommand\prl{Phys. Rev.~Lett.}    
\newcommand\physrep{Phys.~Rep.}      

\usepackage{amssymb}
\usepackage{amsmath}
\usepackage{natbib}
\usepackage{subcaption}
\usepackage[colorlinks=True, citecolor=red]{hyperref}

\usepackage{etoolbox}
\makeatletter
\patchcmd\@combinedblfloats{\box\@outputbox}{\unvbox\@outputbox}{}{%
  \errmessage{\noexpand\@combinedblfloats could not be patched}%
}%
\makeatother

\usepackage{xcolor}

\journal{New Astronomy Reviews}

\begin{document}

\def\alfven{Alfv\'en}
\begin{frontmatter}



\title{Particle acceleration in astrophysical jets}


\author{James H. Matthews$^{1}$, Anthony R. Bell$^2$ \& Katherine M. Blundell$^3$}

\address{$^1$Institute of Astronomy, University of Cambridge, Madingley Road, Cambridge, CB3 0HA\\
$^2$Clarendon Laboratory, Parks Road, University of Oxford, Oxford, OX1 3PU\\
$^3$Astrophysics, Keble Road, University of Oxford, Oxford, OX1 3RH\\
}

\begin{abstract}
In this chapter, we review some features of particle acceleration in astrophysical jets. We begin by describing four observational results relating to the topic, with particular emphasis on jets in active galactic nuclei and parallels between different sources. We then discuss the ways in which particles can be accelerated to high energies in magnetised plasmas, focusing mainly on shock acceleration, second-order Fermi and magnetic reconnection; in the process, we attempt to shed some light on the basic conditions that must be met by any mechanism for the various observational constraints to be satisfied. We describe the limiting factors for the maximum particle energy and briefly discuss multimessenger signals from neutrinos and ultrahigh energy cosmic rays, before describing the journey of jet plasma from jet launch to cocoon with reference to the different acceleration mechanisms. We conclude with some general comments on the future outlook. 
\end{abstract}


\end{frontmatter}


\section{Introduction}
\label{sec:intro}
High-energy, superthermal particles are present in a wide variety of astrophysical systems, particularly those that produce collimated jets of plasma. Indeed, most of what we know about jets comes from observations in the radio, X-ray and gamma-ray bands where relativistic, nonthermal particles are undergoing synchrotron and inverse Compton radiative losses via interactions with either magnetic fields or ambient radiation fields. A significant fraction of the jet's energy budget can be converted into nonthermal particles and the radiation they emit. Remarkably, power-law spectra with cut-offs at various frequencies are observed in jetted systems that have central masses ranging from a few $M_\odot$ up to $\sim10^{10}~M_\odot$. This behaviour hints at, but does not necessitate, universal acceleration physics. 

Astrophysical jets are ideal sites for particle acceleration. They are often supersonic, so can produce strong shock waves. They create environments where velocity shear and turbulence are important, and also transport strong magnetic fields. Each of these effects can lead to the acceleration of high-energy particles in magnetic fields, where particles are accelerated by scattering in a turbulent velocity field or across a shock, shear layer or reconnection site. As we shall see, regardless of the mechanism, the energy increase normally results from moving charged particles through a $-\vec{u} \times \vec{B}$ electric field. The consequent energy gain can be balanced with particle escape from the acceleration site in such a way that a power-law spectrum of particle energies is produced, although this requires special circumstances.

This review is principally focused on how particles can be accelerated to high energies in the jets themselves, the shocks they produce and the turbulence they initiate. We discuss Fermi mechanisms in shocks, turbulence and sites of magnetic reconnection, with reference to other processes. The dominant acceleration process can vary both {\em between} and {\em within} each astrophysical system concerned, and in many cases the detailed acceleration physics is still debated. We generally avoid discussion of detailed observational characteristics except insofar that they inform our physical understanding of particle acceleration in jets. Instead, we refer the reader to the relevant review chapters on respective source types within this volume.

We begin (Section 2) by describing four key observational signatures of high-energy particle acceleration that a successful theory must explain. Section 3 contains the main theoretical discussion, outlining the basic physical principles behind particle acceleration to high energy. In Section 4, we discuss what constrains the maximum energy, before exploring what cosmic ray and neutrino data can tell us about jets and their related phenomena, in Section 5. We discuss our results in Section 6 by considering the journey of plasma along the jet length. Finally, we conclude in Section 7. We use centimetre-gram-second units with particle energies in eV, $u$ and $\Gamma$ to denote bulk velocities and Lorentz factors, $v_{p/e}$ and $\gamma_{p/e}$  to denote particle/electron velocities and Lorentz factors, $E$ for energies and $\vec{{\cal E}}$ for electric field. We use the phrase cosmic ray (CR) to refer to nonthermal protons and ions.

\section{Observations}
\label{sec:obs}
We pick four key observational aspects of astrophysical jets: Images, synchrotron spectra, gamma-ray emission and time-resolved properties. We focus here on electromagnetic radiation and discuss CR and neutrino signals in Section~\ref{sec:multimessenger}. We will use these radiative signatures to inform our discussion of the physics in the next Section.

\subsection{Images}
\label{sec:images}
Images of jetted sources provide important clues about particle acceleration. In radio galaxies, jets can extend $100$s of kpc from the central active galactic nucleus (AGN). The morphologies of the sources are, broadly speaking, split into two \cite{fanaroff_morphology_1974} classes, denoted FRI and FRII. In the FRI sources, the synchrotron emission is brightest near the jet core, with the flux dropping outwards as the jet is disrupted, forming an extended plume of radio-emitting plasma. In the higher power FRII sources, the jet instead remains well-confined for much longer distances and deposits its energy in a termination shock. As with most dichotomies in astrophysics, the FRI-FRII distinction is blurred and a diverse continuum of properties exists that depends on the jet power, composition and environment \cite[e.g.][]{mingo2019}; however, in general, the dependence of FRI/FRII morpology on the radio luminosity is a robust one \citep{fanaroff_morphology_1974,owen_fri/il_1994}.

\begin{figure*}
    \centering
    \includegraphics[width=0.95\linewidth]{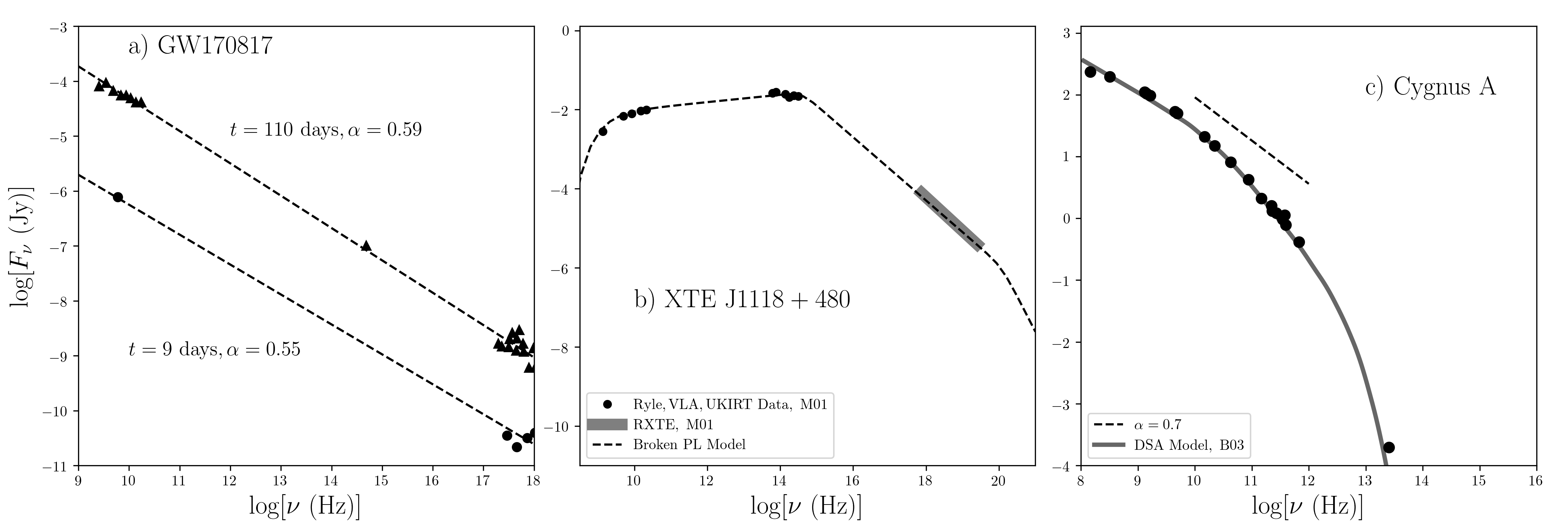}
    \caption{ a)  Broadband spectrum at two different epochs of the afterglow associated with neutron-star merger and short GRB event GW170817, from
    \cite{margutti_binary_2018}. b) Broadband spectrum of the X-ray binary XTE~J1118$+$480 in its low-hard state, compared with an illustrative broken power-law spectrum with $\alpha_{\mathrm{thin}}=0.8$, $\alpha_{\mathrm{thick}}= -0.1$, synchrotron cooling break at $1.2\times10^{20}$~Hz and a low-frequency exponential cut-off at $1.2\times10^{9}$~Hz \citep[][M01]{hynes_x-ray_2000,markoff_jet_2001}. 
    c) Radio spectrum of the southwestern hotspot of Cygnus A \citep{carilli_high-resolution_1999}, with the DSA model from \citet[][B03]{brunetti_-situ_2003} marked in grey. The dashed line shows a power-law with $\alpha=0.7$. 
    The data are digitised from the respective publications. 
    }
    \label{fig:spectra}
\end{figure*}

Radio galaxies produce nonthermal emission at a few distinct locations: compact core regions, the jet beam, lobes, plumes and, in FRIIs, hotspots. Very long baseline interferometry has allowed the jet of M87 (Virgo A) to be imaged in the radio on scales as small as 10 gravitational radii \citep{asada_structure_2012,hada_high-sensitivity_2016,walker_structure_2018}, revealing an edge-brightened morphology with an apparent transition from parabolic to conical streamlines. A similar structure is observed in 3C 84 (Perseus A) at larger distances (100s of gravitational radii) from the central engine \citep{Giovannini2018}. On kiloparsec scales, multi-wavelength images at optical, radio and X-ray wavelengths commonly consist of knots or localised bright spots that imply acceleration in special sites such as shocks \citep{bridle_deep_1994,kraft_chandra_2002,perlman2011,cara2013,clautice2016,perlman2019,wykes_internal_2015,2019ApJ...871..248S}, mixed with smoother emission that suggests distributed particle acceleration \citep{jester_hst_2001,hardcastle_chandra_2005,hardcastle_new_2007,clautice2016,hardcastle_deep_2016,perlman2019}. If the X-ray emission is synchrotron in origin, the short cooling time requires an {\sl in situ} acceleration process \citep{2001ApJ...551..206P,hardcastle_radio_2003,jester2006}, which is generally in-keeping with optical data \citep{jester_hst_2001}. The hotspots are even more robust particle acceleration sites, based on the comparison between the synchrotron cooling time and light travel time, and are consistent with acceleration at the `working surface' of a strong shock \citep[e.g.][]{blandford_twin-exhaust_1974,scheuer_models_1974,meisenheimer_particle_1986,heavens_particle_1987,laing_radio_1989,meisenheimer_synchrotron_1997,brunetti_-situ_2003,araudo_particle_2015}. In FRIs, there is no bright hotspot, but the particle acceleration may still occur primarily while the jet is supersonic and relativistic \citep{laing_systematic_2014}. In most FRIs, the jet is then disrupted and forms a plume-like structure of synchrotron emitting plasma, whereas higher power lobed-FRIs \citep{laing_relativistic_2012} and FRIIs create cocoons and backflows surrounding the jet and extending backwards from the hotspot \citep{hargrave_observations_1974,alexander_ageing_1987,alexander_study_1987,carilli_multifrequency_1991,carilli_cygnus_1996}.

Although radio galaxies are perhaps the most spectacular examples of lobes and hotspots, X-ray binary systems produce similar structures \citep{mirabel_double-sided_1992,mirabel_sources_1999,fender_powerful_2001,corbel_large-scale_2002,blundell_symmetry_2004,jeffrey_fast_2016}, often extending several parsecs into the interstellar medium. X-ray binary jet morphologies change depending on whether the jet is steady or transient \citep{fender_jets_2006}. Transient jets consist of distinct ejections of blobs or plasmoids that can show superluminal motion \citep[e.g.][]{mirabel_notitle_1994,mirabel_sources_1999,miller-jones_notitle_2009,bright_extremely_2020}, while corkscrew-like precession is observed in the slower-moving SS433 jet \citep[e.g.][]{blundell_symmetry_2004,blundell_ss433s_2018}. Protostellar systems such as the Herbig-Haro (HH) objects \citep{herbig_spectra_1951,haro_herbigs_1952} also produce lobes, hotspots and jet emission. Much of this morphology is actually seen in the optical through, e.g., H$\alpha$ or [O\textsc{iii}] emission from shock-heated regions \citep[e.g.][]{lada_cold_1985,reipurth_50_1997,reipurth_herbig-haro_2001}, but radio observations of protostellar jets and HH objects do also reveal signatures of nonthermal emission from high-energy electrons \citep[e.g.][]{carrasco-gonzalez_magnetized_2010,ainsworth_tentative_2014,masque_proper_2015,anglada_radio_2018,rodriguez-kamenetzky_particle_2019}. In some cases, these jets show striking similarities to those in AGN  \citep{carrasco-gonzalez_magnetized_2010}. 

Overall, images of jets tell us that in some sources there are particles being accelerated at the `working surface' of a strong shock, but we must also be able to explain emission from (i) compact regions close to the jet launching point; (ii) both smooth and knotty emission along the jet beam and (iii) more extended plumes, lobes and cocoons clearly associated with jet activity. However, the emission from lobes and cocoons can often be explained by hydrodynamic effects, whereas hotspots, core regions and certain portions of the jet beam appear to be privileged sites for {\sl in-situ} particle acceleration.\footnote{A more complete summary of the observational characteristics of radio galaxy, X-ray binary and protostellar jets can be found in Chapters 3,5 and 10 of this anthology, respectively.}

\subsection{Synchrotron power-laws}
\label{sec:synchrotron}
Perhaps the most universal observational signature of astrophysical jets is synchrotron radiation produced by relativistic electrons spiralling in magnetic fields. The exact spectrum observed depends on the distribution of electron energies, the observing waveband, the number of different radiating electron populations, the bulk Lorentz factor of the radiating plasma and whether the emission is subject to absorption effects. There is also significant diversity in spectral indices and degree of curvature; however, in general, power-law synchrotron spectra with spectral indices $\alpha\approx0-1$ (where $\alpha$ is defined such that $F_\nu \propto \nu^{-\alpha}$, where $F_\nu$ is the flux density in erg~s$^{-1}$~cm$^{-2}$~Hz$^{-1}$) are common in jetted systems. 

To illustrate this behaviour, broadband spectra for three systems with jets are shown in Fig.~\ref{fig:spectra}. In the GRB afterglow from GW170817, an optically synchrotron power-law with $\alpha \approx 0.55$ extends from radio through to X-ray \citep{margutti_binary_2018}. In the low-hard state observations of X-ray binary XTE~J1118$+$480, the spectrum is characterised by a broken power-law and is only optically thin at IR/optical through X-ray frequencies \citep{hynes_x-ray_2000,markoff_jet_2001}. In the radio galaxy Cygnus A, the hotspot shows an optically thin power-law \citep{meisenheimer_synchrotron_1997,carilli_high-resolution_1999,brunetti_-situ_2003}, with hints of a low-frequency cutoff that has been confirmed with {\sl LOFAR} observations \citep{mckean_lofar_2016}. The spectrum exhibits curvature and deviates from a pure power-law which may be due to synchrotron cooling or nonlinear acceleration effects (the latter is discussed in Section~\ref{sec:nonlinearity}). We also show a broadband spectral energy distribution (SED) for the BL Lac type blazar, Mrk 421, in Fig.~\ref{fig:humps}, in which a synchrotron spectrum is observed from radio to X-ray frequencies (the gamma-rays are discussed in the next subsection). 

One way of parameterising an electron population subject to synchrotron cooling losses is via a broken power-law \citep[e.g.][]{heavens_particle_1987,rybicki-lightman,longair_high_1994}, in which the spectrum steepens at high frequencies due to synchrotron losses and the low-frequency emission is self-absorbed. In this case, the spectral shape is given by
\[
    F_{\nu} \propto
\begin{cases}
    \nu^{-\alpha_{\mathrm{thick}}},& \nu < \nu_0, \mathrm{Optically~thick} \\
    \nu^{-\alpha_{\mathrm{thin}}},& \nu_0 < \nu < \nu_\mathrm{cool}, \mathrm{Optically~thin} \\
    \nu^{-\alpha_{\mathrm{thin}}-1/2},& \nu > \nu_\mathrm{cool} , \mathrm{Cooled}\\
\end{cases}
\]
where $\nu_0$ and $\nu_\mathrm{cool}$ denote the frequencies above which the emission becomes optically thin and the emitting electrons' synchrotron cooling timescale is less than the time since the electron was accelerated, respectively. In reality, model spectra show a gradual transition rather than sharp breaks \citep[e.g][]{jaffe-perola_1973,heavens_particle_1987,brunetti_-situ_2003}. At low frequencies self-absorption produces either an inverted spectrum ($\alpha_{\mathrm{thick}}\approx-5/2$; in the case of a single self-absorbed population of electrons) or a flat-spectrum with $\alpha_{\mathrm{thick}} \sim 0$ \citep[in the case of a conical self-absorbed jet; ][]{blandford_relativistic_1979}. At intermediate frequencies ($\nu_0 < \nu < \nu_c$), we typically observe $\alpha_\mathrm{thin} \approx 0.5-0.8$. If this optically thin emission comes from a single population of electrons with a constant magnetic field $B$ and power-law distribution of particle energies $n(E) dE \propto E^{-s} dE$, then the relationship between $s$ and $\alpha$ is given by $\alpha_\mathrm{thin} = (s-1)/2$, implying $s\approx2-2.6$. 

The characteristic emission frequency of an electron with Lorentz factor $\gamma_e$ spiralling in a magnetic field $B$ is given by  $\nu_c = \Gamma \gamma_e^2 eB / (2 \pi m_e c)$, where $\Gamma$ is the bulk Lorentz factor of the flow. Since astrophysical jets can be highly relativistic, the critical frequencies ($\nu_0$, $\nu_\mathrm{cool}$,...) can be boosted by large factors if the jet is beamed towards us. The low-frequency cutoff is determined by either the minimum Lorentz factor of the electrons ($\gamma_m$) multiplied by $\Gamma$, or absorption effects \citep[e.g.][]{mckean_lofar_2016}. In some cases, such as at early times in GRBs, the `fast-cooling' limit is realised and $\nu_c(\gamma_m) > \nu_{\mathrm{cool}}$ \citep{sari_spectra_1998,granot_synchrotron_2000}. Steepening at high-frequency can be caused by synchrotron or inverse Compton cooling and, in GRBs in particular, the time evolution of this break provides useful information about particle populations; however, it has been shown that in radio galaxies the high-frequency cutoff can reflect the maximum particle energy rather than cooling effects \citep[see e.g. Section~\ref{sec:max_energy},][]{araudo_evidence_2016,araudo_maximum_2018}. The break frequency between optically thick and optically thin emission ($\nu_0$) can also be used to infer information about jet dynamics and sites of particle acceleration, in, e.g. X-ray binary jets \citep{markoff_jet_2001,gandhi_variable_2011,koljonen2015}.

\subsection{Gamma-ray emission}
\begin{figure}
    \centering
    \includegraphics[width=1.0\linewidth]{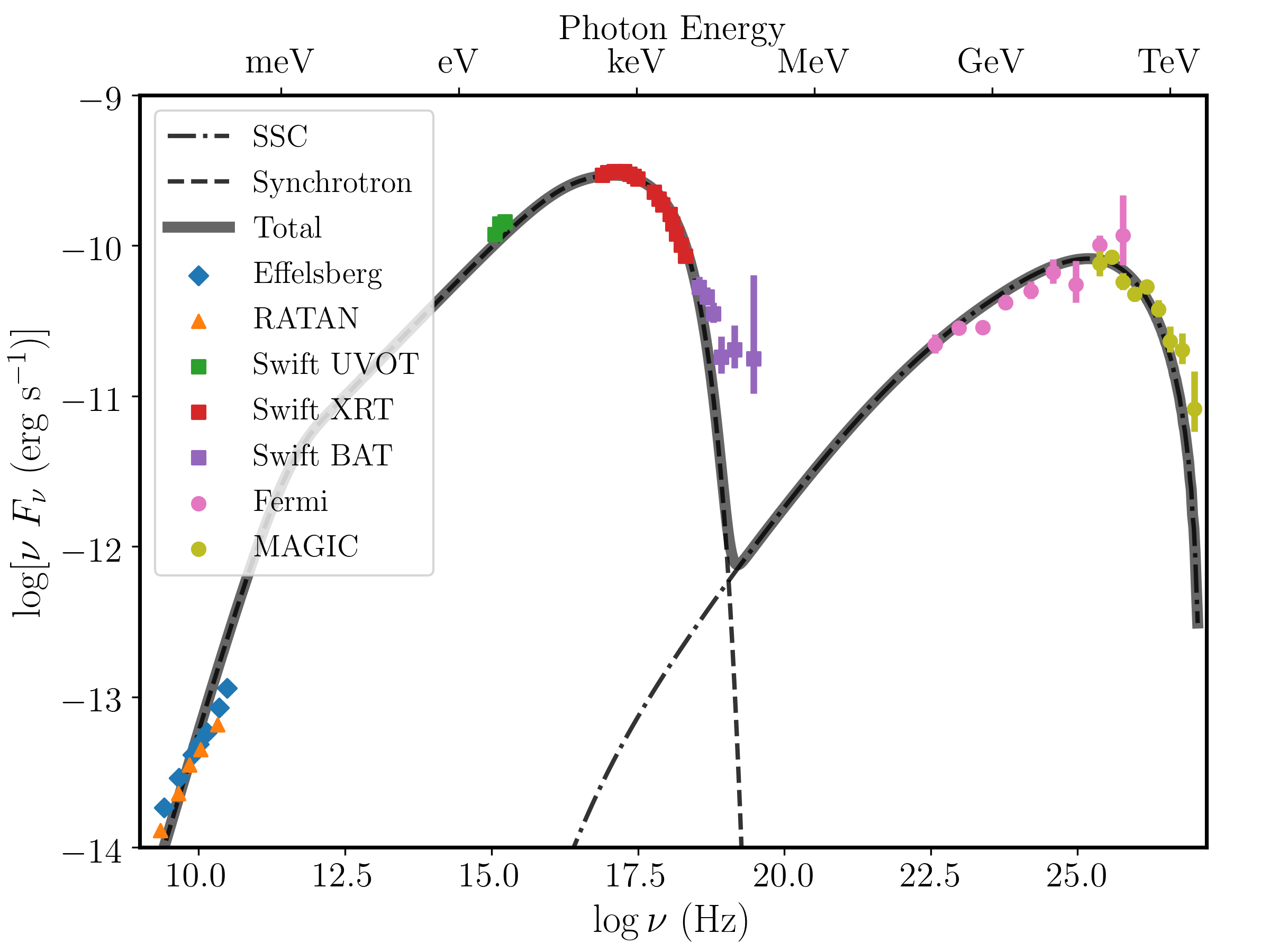}
    \caption{Broadband SED of the BL Lac class blazar Mrk 421 from the multi-wavelength campaign of \cite{abdo_fermi_2011}. The different data sources are labelled and in different colours. Also shown, for illustrative purposes, is a one-zone synchrotron plus SSC model, produced with \textsc{naima} \citep{zabalza_naima:_2015}. The model has $B=0.08$G and uses an electron distribution with a broken power-law with indices $2.2$ and $2.7$ with a break at $\gamma_e=5\times10^{4}$ and an exponential cutoff at $\gamma_e=5\times10^{5}$. The spectrum shows the classic `double-humped' shape that is typical of blazars and other high-energy astrophysical particle accelerators. 
    }
    \label{fig:humps}
\end{figure}

Gamma-ray emission has been observed in a wide variety of jetted systems, with examples including GRB prompt emission \citep[e.g.][]{klebesadel_grbs_1973,gehrels_gamma-ray_2009}, microquasars such as SS433 and Cygnus X-3 \citep{araudo_transient_2011,corbel_giant_2012,bordas_detection_2015,abeysekara_very-high-energy_2018}, radio galaxies such as Centaurus A and Fornax A \citep{abdo_fermi_2010,abdo_fermi_2010-1,ackermann_fermi_2016} and blazars\footnote{See Chapter 4.}  \citep{ghisellini_general_2010,acciari_tev_2011,abdo_fermi_2011,tavecchio_gamma_2017}. In most of the sources where broadband spectral information exists, the SED shows a distinctive `double hump' shape characterised by a low-energy synchrotron spectrum and high-energy bump. Fig.~\ref{fig:humps} shows an example for the well-studied blazar Mrk 421 from the observing campaign of \cite{abdo_fermi_2011}. This double-humped spectrum is observed across the blazar sequence \citep{fossati_unifying_1998}. The low-frequency hump is well-established as synchrotron emission, but the origin of the high-frequency emission is less certain \citep[e.g.][]{blandford_relativistic_2018, biteau_progress_2020}.

Gamma-rays can be produced by both {\em leptonic} and {\em hadronic} interactions. In the leptonic case, the radiation is due to the inverse Compton (IC) interactions of energetic electrons with a radiation field such as the CMB, thermal emission from the accretion process, or the electrons' own synchrotron radiation \citep[e.g.][]{dermer_model_1993}. In the latter case, this is referred to as synchrotron self-Compton (SSC). An illustrative synchrotron plus SSC model is shown in Fig.~\ref{fig:humps}. One-zone formulations of synchrotron-Compton models are commonly invoked to fit blazar spectra \citep[e.g.][]{maraschi_jet_1992,kirk_particle_1998,ghisellini_general_2010}, but these models can encounter problems when fitting the spectral shape \citep[e.g.][]{bednarek_testing_1997}, or by requiring that the synchrotron cooling time is longer than the variability timescale \citep[e.g.][]{balakovic2016}. However, more complete treatments in which the electron population is self-consistently evolved along a model jet can provide good fits to blazar spectra  \citep{potter_synchrotron_2012,potter_synchrotron_2013,potter_synchrotron_2013-1,morris_feasibility_2018}. Other leptonic models have been proposed for gamma-ray production, such as $e^+e^-$ pair cascades \citep{blandford_pair_1995,bednarek_gamma-rays_1997}. In hadronic models, gamma-rays are produced when high-energy protons interact with either other protons, photons or magnetic fields, with the relative importance of each depending on the density, radiation energy density and magnetic field energy density \citep[e.g.][]{mannheim_proton_1993,mucke_bl_2003,de_angelis_2018}. In $p\gamma$ interactions, neutral pions are produced via $p\gamma \rightarrow p \pi^-$, and pairs are produced via the Bethe-Heitler process, $p\gamma \rightarrow p e^+ e^-$; subsequent pion decay ($\pi_0\rightarrow \gamma \gamma$)  or annihilation then produces gamma-rays. Inelastic $pp$ collisions also create neutral pions which decay in the same manner. Each of the pion production channels also leads to charged pions, which ultimately create neutrinos, meaning hadronic gamma-rays are intrinsically linked to high-energy neutrinos (see Section~\ref{sec:multimessenger}). Finally, protons can produce synchrotron radiation when the magnetic field is strong enough \citep{aharonian_tev_2000,aharonian_constraints_2002,biteau_progress_2020}. 

Distinguishing between hadronic and leptonic gamma-ray emission is challenging, but, at least in principle,  gamma-rays offer one of the few possible ways to study the composition of jets on small scales, particularly in GRBs and blazars. Constraining jet composition is important in order to (i) help differentiate between launching mechanisms (for example, whether the angular momentum extracted originates from a BH ergosphere [\citealt{blandford_electromagnetic_1977}] or accretion disc [\citealt{blandford_hydromagnetic_1982}]); (ii) understand multimessenger signals from UHECRs/neutrinos (see Section~\ref{sec:multimessenger}) and (iii) accurately estimate the energy partitioning in the jet and resulting effect on surroundings.  

\subsection{Variable sources and transients}
We can also use the time domain to learn about particle acceleration. In GRBs or X-ray binaries, where dynamical timescales are short compared to those in AGN, time-resolved observations can be used to produce light-curves over the full lifetime of the outburst or explosion. However, short-timescale changes can also occur in AGN, examples being blazar flares \citep{aharonian_exceptional_2007}, jet variability in e.g. M87 \citep{giannios_fast_2010} and Pictor A \citep{marshall_flare_2010}, and even variability in radio galaxy hotspots \citep{hardcastle_chandra_2005}.

Blazars are split into two main classes, BL Lacs and flat-spectrum radio quasars (FSRQs). Similarly to FRI/FRII radio galaxies, these classes can be divded approximately by their jet power \citep{celotti_power_2008}, with FSRQs the more powerful population. Extremely fast variability has been detected in both classes of blazar, with $\sim10$~minute flares observed at TeV energies in, e.g., the BL Lac PKS 2155-304 \citep{aharonian_exceptional_2007} and at GeV energies in, e.g., the FSRQ 3C 279 \citep{ackermann_minute-timescale_2016}. This behaviour requires that any complete model for particle acceleration can produce rapid variability in some parts of the blazar jet. 

Intriguing timing properties are also seen in X-ray binaries. During a recent outburst of V404 Cygni, \cite{gandhi_elevation_2017} found $0.1$~second lags between the X-ray and optical variability, associating this time lag with the distance from the black hole to the optical-emitting region in the jet. In GRS 1915$+$105, \cite{de_gouveia_dal_pino_production_2005} suggested that the superliminal motion seen in the jet \citep{mirabel_notitle_1994,mirabel_sources_1999} can be explained by fast reconnection occurring just above the black hole, which can cause rapid X-ray variability. In both V404 Cygni and GRS 1915$+$105, the models invoked draw upon blazar phenomenology and theory in an attempt to unify some of the physical mechanisms at work in both X-ray binaries and AGN. 

Time-resolved observations of GRBs\footnote{See Chapter 7.}  reveal two main populations: short-duration GRBs, which were recently confirmed to be associated with a binary neutron star merger \citep{abbott_gravitational_2017}; and long-duration GRBs, thought to be caused by collapsars \citep{galama_unusual_1998,iwamoto_hypernova_1998,macfadyen_collapsars:_1999}. The classes are split by the length of the prompt emission, with the dividing line at $\approx 2$s \citep{kouveliotou_identification_1993}. Following the prompt emission, there is a longer-lived `afterglow' phase of synchrotron emission from a shock front \citep[e.g.][]{waxman_-ray_1997,sari_spectra_1998,piran_gamma-ray_1999,yost_study_2003,kumar_physics_2015}. The evolution of the spectrum can be used to infer information about the particle distribution function and magnetic field \citep{sari_spectra_1998,wijers_physical_1999}, while breaks in the power-law slopes of the light curves provide useful constraints on the jet dynamics and geometry \citep{rhoads_how_1997,sari_jets_1999,kumar_steepening_2000,de_colle_simulations_2012}.
X-ray binary sources also show characteristic rise and decay phases \citep[e.g.][]{fender1999,fender_transient_2006,corbel_giant_2012,kim_flaring_2016}, with the morphology and spectrum of the radio emission depending on the accretion state. In hard or quiescent states, optically thick emission from a compact jet is observed \citep{fender_powerful_2001,fender_towards_2004,gallo2005,russell_jet_2013}, with $\alpha \lesssim 0$, and the transient ejections described in section~\ref{sec:images} are then produced as the source crosses into the soft state \citep{mirabel_sources_1999,fender_towards_2004} often in combination with an optically thin radio flare. Discrete ejection events can be modelled using the `synchrotron bubble' model \citep{van_der_laan_model_1966},
or variants thereof \citep{hjellming_radio_1988,tetarenko2017,tetarenko_tracking_2019}, but it can be necessary to account for more complex time-dependent spectral evolution \citep[e.g.][]{atoyan1999,miller-jones_time-sequenced_2004}. While the variability is often primarily driven by accretion physics and jet dynamics, good models for the acceleration and subsequent evolution of the synchrotron electrons are important.

\section{Theory of Particle Acceleration}
\label{sec:theory}
There already exist a number of reviews covering the theory of particle acceleration; some relevant examples include those focusing on
shock acceleration \citep{drury_introduction_1983,blandford_particle_1987,schure_diffusive_2012,bell_particle_2014,marcowith_microphysics_2016} and magnetic reconnection
\citep{hoshino_relativistic_2012,de_gouveia_dal_pino_particle_2015,kagan_relativistic_2015}, as well as more general discussions of astrophysical jets
\citep{begelman_theory_1984,mirabel_sources_1999,fender_jets_2006,bykov_particle_2012,romero_relativistic_2017,blandford_relativistic_2018}, particle acceleration simulations \citep{marcowith2020} or high-energy astrophysics \citep{longair_high_1994}. For this reason, we only cover some of the essential theory and refer the enthusiastic reader to these efforts. Much of the physics is quite general and not necessarily specific to jets. 

We begin by writing the equation for the rate of change in momentum of a particle with Lorentz factor $\gamma$ and velocity $\vec{v}_p$: 
\begin{equation}
    \frac{d}{dt} (\gamma m \vec{v}_p) = 
    e (\vec{{\cal E}} + \vec{v}_p \times \vec{B}),
\end{equation}
where $\vec{{\cal E}}$ is the electric field and $\vec{B}$ the magnetic field. The motion of free charges means that large electrostatic fields are difficult to maintain in the mostly ionized media that are common in astrophysical systems, so the electric field is often just $-\vec{u} \times \vec{B}$, where $\vec{u}$ is now the {\em bulk} velocity of the plasma. It is this $-\vec{u} \times \vec{B}$ term that leads to particle acceleration. Given the magnetic nature of the acceleration, it is also useful to define the Larmor radius (or gyroradius) of the particle, $r_g=p/(ZeB)$, where $Ze$ is the charge on the particle and $p$ the momentum. For relativistic particles with energy $E=pc$ this becomes
\begin{equation}
    r_g = \frac{E}{ZecB}.
\label{eq:confinement}
\end{equation}
This gives the first constraint on acceleration mechanisms: the acceleration site must have a size $R$ larger than $r_g$. We call this the {\em confinement condition}. 

Let us now consider a particle undergoing some form of scattering process in which it gains energy. We assume we have $N_0$ particles in the process, with initial energy $E_0$. Each time the particles undergo a collision, they gain a fractional increase in energy of $\beta_i$ such that after $m$ collisions they have energy $E=E_0 \prod_{i=1}^m \beta_i$. Similarly, if they have a probability $P_i$ of remaining in the acceleration region after each collision, then the number of particles left after $m$ collisions, with energy $E$, is $N = N_0 \prod_{i=1}^m P_i$. We can therefore write 
\begin{equation}
    \frac{\ln(N/N_0)}{\ln(E/E_0)} =  \frac{\sum_{i=1}^{m} \ln P_i}{\sum_{i=1}^{m} \ln \beta_i}.
\end{equation}
Observed synchrotron spectra (Section~\ref{sec:synchrotron}) and the cosmic ray spectrum (Section~\ref{sec:multimessenger}) require a powerlaw distribution for the differential number of particles $n(E)=dN/dE$. A power-law distribution for $n(E)$ is only possible when $(\sum_{i=1}^{m} \ln P_i) / (\sum_{i=1}^{m} \ln \beta_i)$ is constant regardless of the number of collisions. One way of achieving this is if $P_i$ and $\beta_i$ are constant values with increasing particle energy, which we label $P$ and $\beta$. In this special case, one obtains a power-law of the form 
\begin{equation}
    n(E)~dE \propto E^{(\ln P/\ln \beta)-1} dE.
    \label{eq:energy_spec}
\end{equation}
This equation shows that particles undergoing stochastic collisions can produce a power-law distribution of particle energies whose spectral index depends on the energy gain per collision and escape probability, but only if $P_i$ and $\beta_i$ remain constant with increasing energy. The fractional energy gain might feasibly be roughly constant with increasing energy, but arranging for $P_i$ to be independent of energy is difficult. The escape probability per collision $(1-P_i)$ is the ratio of the scattering time (the interval between scattering events), $\Delta t$, to the escape time, $\tau_{\mathrm{esc}}$. The scattering time is $\Delta t \sim \lambda / c \propto r_g/c$, where $\lambda$ is the mean free path. If the escape is diffusive, then $\tau_{\mathrm{esc}} \propto r_g^{-1}$, which then implies $(1-P_i) \propto r_g^{2} \propto E^{2}$.  As we shall see, one way of getting around this apparent difficulty is via shock acceleration, which provides a physically motivated reason for $P_i$ to be constant with increasing energy. Acceleration can also come from other electric fields and non-stochastic processes (see in particular Section~\ref{sec:reconnection}). Nonetheless, this simple model with escape balancing energy gain provides a useful framework that aids discussion of a variety of different acceleration mechanisms. 

\subsection{Second-order Fermi Acceleration}
In second-order Fermi acceleration \citep[Fermi II;][]{fermi_origin_1949}, particles scatter off magnetised clouds or in turbulent velocity field with random velocities. This process can be understood by considering a series of `magnetic mirrors' that either reflect the particle or isotropise the particle distribution function in the rest frame of the scattering centre. In Fermi's approach, the scatterers are `clouds' moving with velocity $u_c$, and relativistic particles gain or lose a fraction $\sim u_c/c$ of their energy per collision. The energy increase comes from the fact that head-on collisions are more likely. For relativistic test particles, the average energy gain per collision is found by tranforming from the lab frame to the cloud frame and back, then integrating over the pitch angle distribution \citep[e.g.][]{longair_high_1994}, and is given by
\begin{equation}
    \left \langle \frac{\Delta E}{E} \right \rangle = \frac{8}{3} \left(\frac{u_c}{c} \right)^2 .
\end{equation}
Alternatively, if the energy gain per collision is small, Fermi acceleration can be treated using the Fokker-Planck equation \citep[e.g.][]{blandford_particle_1987}. Although Fermi's original discussion considers clouds, subsequent elaborations \citep[e.g.][]{kulsrud_relativistic_1971,melrose_plasma_1980} have instead considered scattering off MHD waves, in which case the relevant velocity becomes the \alfven\ speed, $v_A = B/\sqrt{4\pi\rho}$. The acceleration rate for Fermi II is second-order in $(u_c / c)$ or $(v_A / c)$. As a result, Fermi II is slow and a particle must be confined for a long time if the energy is to reach a large value, except in special conditions \citep[e.g.][]{jones1994}. Furthermore, as discussed above, there is no obvious reason why the escape time from a volume of turbulent plasma should scale with the scattering time so as to achieve a constant escape probability -- in fact, the opposite behaviour ($\Delta t \propto 1/\tau_\mathrm{esc}$) is expected if the escape is diffusive. Even if this can be accomplished, there is then no {\em a priori} reason why a given spectral index should be produced. As a result, a Fermi II model requires significant fine-tuning in order to produce the observed power law spectra and their spectral index values.

\subsection{Shock Acceleration}

\begin{figure*}
    \centering
    \includegraphics[width=0.8\linewidth]{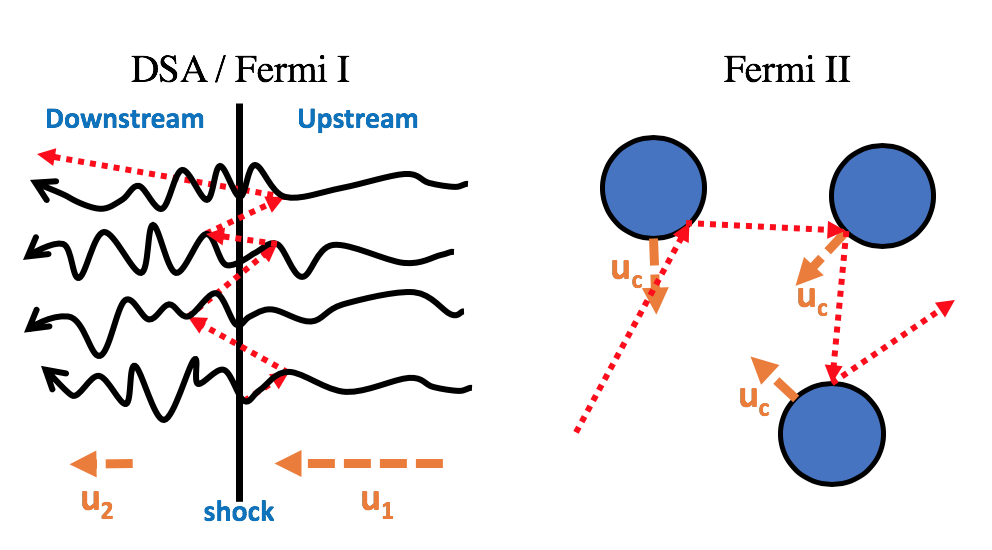}
    \caption{Diagram showing, schematically, how particles are accelerated in DSA (left) and in Fermi II (right). Bulk velocities are shown with orange dashed lines, crude particle trajectories in red dashed lines and magnetic field lines as black arrows. In the DSA case, the velocities are in the shock rest frame.
    }
    \label{fig:shock}
\end{figure*}

Particles can be accelerated by a few different mechanisms at collisionless shock waves: shock drift acceleration \citep[SDA; ][]{begelman_shock-drift_1990,decker_shock_1985,chalov_shock_2001}, shock-surfing acceleration \citep[SSA;][]{sagdeev_cooperative_1966,katsouleas_unlimited_1983,lee_pickup_1996,zank_``injection_2001,shapiro_shock_2003} and diffusive shock acceleration (DSA, described below). These mechanisms and the relevant microphysics is summarised by \cite{marcowith_microphysics_2016}. SDA occurs in quasi-perpendicular shocks and involves the drifting of particles along the shock front; the drifting is effectively a $\nabla B$ drift introduced by the compression of the magnetic field across the shock front that leads to tighter gyrations in the downstream field with respect to the upstream. SSA is similar, and occurs when ions are `reflected' by the potential of the shock and then return from upstream. SSA can only accelerate particles to mildly suprathermal energies since the particle quickly becomes energetic enough to overcome the shock potential. A single episode of SDA also normally leads to only a small energy gain because the particle drifts away from the shock. To reach highly superthermal energies in a quasi-perpendicular shock a particle must return to the shock multiple times and undergo repeated episodes of SDA. This version of first-order Fermi acceleration is thus similar to convential DSA; the difference between SDA and DSA is in many ways a superficial and semantic one, and first-order Fermi acceleration at shocks does not need to involve only drifts or only diffusion \citep[e.g.][]{jokipii_particle_1982}. We discuss drifts and SSA further in the context of injection in Section~\ref{sec:injection}. In this section, we discuss the general features of first-order Fermi acceleration at shocks using a DSA framework. 

The theory of DSA was laid out in four landmark papers \citep{axford_acceleration_1977,krymskii_regular_1977,bell_acceleration_1978,blandford_particle_1978}, while similar mechanisms had earlier been described by \cite{jokipii_model_1966} and \cite{fisk_increases_1971}. The presence of a shock allows for a situation in which `collisions' are essentially always head-on, although the actual scattering process is deflection in a turbulent magnetic field. To illustrate this, let us consider an already relativistic test particle, such that we can set $v_p = c$ and $E=pc$, encountering a non-relativistic, collisionless shock with upstream and downstream velocities $u_1$ and $u_2$ (see Fig.~\ref{fig:shock}). We follow \cite{bell_acceleration_1978} in considering a steady state solution where the number of particles being injected at the shock is equal to the number being carried away downstream, and we assume that all particles from upstream are scattered back towards the shock by self-generated turbulence. 

In this scenario, the particle's energy as it crosses from the upstream to downstream regions is given by a Lorentz transformation such that $E^{\prime} = \gamma_s (E + 3 p u_2 \cos \theta)$, since for a strong shock the downstream plasma has a velocity of $3 u_2$ relative to the upstream plasma. Here $\theta$ is the `pitch angle', the angle with respect to the shock normal. An integration over pitch angles reveals that the fractional energy gain per half cycle is $2 u_2 / c$, and thus the quantity $\ln(\beta)$, which includes a factor of two for the full cycle there and back across the shock, is given by 
\begin{equation}
\ln \beta = \ln \left( 1 + \frac{4 u_2}{c} \right) \sim \frac{u_1}{c}.
\end{equation}
The probability $P$ that the particle remains within the acceleration region can be calculated by considering that, in steady state, the number of particles being removed from the shock by being swept downstream is $Nu_2$, whereas the number of particles crossing the shock is $Nc/4$. Thus the fraction of particles lost from the acceleration site per unit time ($1-P$) is $4Nu_2 / Nc$ and we can write
\begin{equation}
\ln P = \ln \left(1-\frac{4u_2}{c} \right) \sim -\frac{u_1}{c}.
\end{equation}
Crucially, neither the rate at which particles leave the acceleration region nor the fractional energy gain they experience depends on energy. Both are instead constants that depend only on the shock velocity. Combining these two expressions with equation~\ref{eq:energy_spec} then results in  a spectral index of $s=2$. To obtain this spectral index, we have assumed a strong shock (Mach number, $M \rightarrow \infty$) and an adiabatic index of $5/3$ such that the shock compression ratio $\chi=4$, as well as fully relativistic particles, but the more general result is a power-law in momentum in the particle distribution function such that 
\begin{equation}
f(p) \propto p^{-3\chi/(\chi-1)},
\end{equation}
which gives $f(p)\propto p^{-4}$ for $\chi=4$. The relationship to the differential number density is $n(p) = 4 \pi p^2 f(p)$ for an isotropic particle distribution, so $n(p) \propto p^{-2}$.

DSA is attractive as an acceleration mechanism because the energy gain is first-order in shock velocity, and because the balance between escape and fractional energy gain is hardwired by the shock jump conditions; thus, the index $s$ in the power-law energy distribution does not require fine-tuning and there is a good reason why a power-law with a fairly universal spectral index might exist. Since the theory was first developed in the late 1970s, the theoretical understanding of DSA has developed considerably. We therefore discuss considerations that go beyond the simple treatment we have outlined, by considering non-linearity, injection and relativistic shock velocities with reference in particular to recent progress using numerical simulations. 

\subsubsection{Non-linearity and self-generated turbulence}
\label{sec:nonlinearity}
Non-linearity in shock acceleration is introduced via the back-reaction of the high-energy particles on the conditions of the background plasma. This is inevitable if particle acceleration is efficient since a relatively high fraction of the energy density will be contained in superthermal particles. In this case the CR pressure must be included in the fluid equations and this modifies the shock structure \citep{axford_acceleration_1977,drury_hydromagnetic_1981,drury_introduction_1983,drury_stability_1986}. Applying a two-fluid model reveals two main effects: the smoothing of the shock on the diffusion length, and an increase of the compression ratio $\chi$ due to the relativistic CR equation of state (adiabatic index of $4/3$). The CRs can also carry away internal energy from the shock region, effectively acting as a coolant and increasing the compression ratio further. An increased compression ratio acts to flatten the spectrum at high energies \citep[see e.g.][]{ellison_non-linear_1995}.

Non-linearity is also introduced via waves or plasma instabilities excited by CRs. This is a crucial ingredient of DSA theory, since turbulence is needed to scatter the particles back across the shock front, as noticed by \cite{bell_acceleration_1978} and \cite{blandford_particle_1978}. Both originally invoked Alfv\'en waves excited by CRs streaming with super-Alfv\'en velocities ahead of the shock. This effect and its derivatives are often referred to as the resonant, streaming or Alfv\'en instability, so-called because it grows at wavenumbers resonant with the Larmor radius of the streaming CRs \citep{lerche_unstable_1967,kulsrud_effect_1969,wentzel_cosmic-ray_1974,skilling_cosmic_1975,skilling_cosmic_1975-1,skilling_cosmic_1975-2}. The resonant instability saturates at $\delta B/B \sim 1$ \citep{mckenzie1982,amato_kinetic_2009}. 

A key physical limit of diffusive particle transport is that of Bohm diffusion, when the scattering mean free path is roughly equal to the Larmor radius of the particle. In this case, the diffusion coefficient is given by $D_B=r_gc /3$. In supernova remnants (SNRs), there is observational evidence that Bohm diffusion applies \citep{stage_cosmic-ray_2006,uchiyama_extremely_2007}, but this can only be the case if the turbulent field is comparable to the uniform field, i.e. $\delta B/B \sim 1$. As a result, turbulent magnetic field amplification is a general feature of DSA theory and can affect the maximum attainable energy (see Section 4) as well as the spectrum of the accelerated particles \citep{caprioli_2009,bell_steepening,malkov_2019}

In high Mach number shocks, there is an instability that grows faster than the resonant instability at wavenumbers non-resonant with the Larmor radius, known as the non-resonant hybrid (NRH) or Bell instability \citep{lucek_non-linear_2000,bell_turbulent_2004,bell_interaction_2005}. The basic physics of the instability is that a return current $-\vec{j}_{\mathrm{cr}}$ produced in reaction to streaming CRs stretches and distorts the field via a $-\vec{j}_{\mathrm{cr}} \times \vec{B}$ force. Numerical simulations have shown that the instability grows exponentially even in the non-linear regime and amplifies the magnetic field to $\delta B/B \sim 10-100$ \citep[e.g.][]{zirakashvili_modeling_2008,reville_transport_2008,riquelme_nonlinear_2009,riquelme_magnetic_2010,matthews_amplification_2017}, although the saturated field value depends on the particle energy regime considered. \cite{caprioli_2014b} find, using PIC simulations, a transition from resonant to non-resonant instability at an \alfven ic Mach number, $M_A$, of $30$; in their study, the magnetic field is amplified in both regimes with an amplification factor that scales as $\sqrt{M_A}$. 
One important general property of the NRH instability is that the scale-size of the turbulence grows until it reaches the Larmor radius of the particles driving the instability, thus providing a natural way for the Bohm limit to be realised. Although most of the work on the NRH instability has focussed on SNRs, it has also been applied to jets in radio galaxies \citep{araudo_maximum_2018} and GRBs \citep{milosavljevic_cosmic-ray_2006}. 

\subsubsection{Relativistic shocks}
\label{sec:relativistic_shocks}
Given that astrophysical jets are often relativistic, we must consider how the original analysis changes as the shock velocity enters the relativistic regime. One of the main differences is that the energy spectrum is steeper than the non-relativistic case; \cite{kirk_particle_2000} find $s=2.23$ and \cite{achterberg_particle_2001} find $s=2.2-2.3$, while \cite{sironi_maximum_2013} find an even steeper spectrum from particle-in-cell (PIC) simulations. The reason for the steeper spectral index compared to non-relativistic shocks is due to a combination of anisotropy in the particle distribution function and the time available for scattering \citep[see e.g.][for a summary]{bykov_particle_2012}. Although the simple diffusion approximation employed above breaks down, the spectral steepening can be na\"ively understood as due to a smaller value of $\beta$ compared to non-relativistic shocks \citep{achterberg_particle_2001}. In addition, particles in the upstream region do not diffuse far upstream and instead are rapidly overtaken by the shock, before the direction of the particle has been significantly altered, while particles in the downstream region are rapidly advected downstream. Thus, in both the upstream and downstream regions CRs have limited time available to generate a turbulent magnetic field.  

The ability of CRs to generate turbulence is crucial and at very high bulk Lorentz factors this effect can prohibit DSA occuring at all \citep{sironi_relativistic_2014}. At an ultra-relativistic shock ($u_s\rightarrow c$, $\Gamma \gg 1$), the combination of quasi-perpendicular magnetic field, steep CR spectrum and reduced time available before the shock overtakes, or advects away, the CRs means that both the scale-size and magnetic field amplification associated with the turbulence generated by streaming or drifting CRs are severely limited. As a result, diffusion occurs well above the Bohm regime and the maximum particle energy is lowered accordingly \citep[][see also Section 4]{lemoine_electromagnetic_2010,reville_maximum_2014,bell_cosmic-ray_2018}. 

\subsubsection{Injection}
\label{sec:injection}
The DSA theory outlined above assumes that the particle being injected in the system has $v_p \gg u_s$ and is already relativistic -- that is, the {\em injection} of the particle from the thermal pool onto the superthermal power-law tail is not considered. This ``injection problem'' is a quite general issue for Fermi-type processes, although much work has been devoted to understanding the injection into the DSA process in particular \citep[e.g.][]{bell_acceleration_1978-1,giacalone_ion_1993,levinson_injection_1996,gruzinov_gamma-ray_1999,zank_``injection_2001,nishikawa_particle_2003,spitkovsky_particle_2008,caprioli_2014a,marcowith_microphysics_2016}. The injection process is quite different to DSA at the highest-energies and requires a full kinetic treatment that includes the backreaction on the fluid. Injection is therefore a highly-nonlinear problem that is often studied using PIC simulations. Over the past decade or so, PIC and hybrid MHD-PIC simulations have been hugely successful in modelling the DSA process at low energies and self-consistently producing the turbulence that is so critical for its operation \citep[e.g.][]{spitkovsky_particle_2008,sironi_synthetic_2009,sironi_particle_2009,sironi_particle_2011,riquelme_electron_2011,caprioli_2014a,caprioli_2014b,caprioli_simulations_2015,bai_magnetohydrodynamic-particle--cell_2015,crumley_kinetic_2019}. 

A variety of injection mechanisms have been proposed, often involving microinstabilites excited at the shock. For example, the Weibel instability \citep{weibel_spontaneously_1959} can mediate collisionless shocks and lift particles from thermal to suprathermal or mildly superthermal energies 
\citep{spitkovsky_particle_2008}. Other possibilities are that particles are injected by interaction with whistler waves \citep[e.g.][]{baring_particle_1991,riquelme_electron_2011}, or, if the shock is locally quasi-perpendicular, via SDA \citep[e.g.][]{ball_shock_2001,park_simultaneous_2015}. Injection efficiencies can be expected to vary between electrons and ions for non-relativistic shocks since at non-relativistic temperatures the electrons have gyroradii a factor $\sqrt{m_e/m_p}$ smaller than protons and it is harder to scatter them across the shock \citep[e.g.][]{bell_acceleration_1978-1}. PIC simulations do show that protons are injected more efficiently than ions \citep{sironi_particle_2011,park_simultaneous_2015,crumley_kinetic_2019}, with the electron efficiency increasing as the electron temperature $k_B T_e$ approaches $m_e c^2$ \citep{sironi_particle_2011,crumley_kinetic_2019}. Possible observational evidence for this phenomenon can be found by comparing the energy fraction in electrons at non-relativistic supernova shocks, where it is low \citep{morlino_strong_2012}, to relativistic GRB shocks, where it is higher \citep{wijers_physical_1999}. 

Particles may also be injected by being reflected from the shock, due to either the electrostatic shock potential or magnetic mirror effects at oblique shocks \citep{hudson_reflection_1965}. These ions subsequently return from upstream with a fractional energy gain on the order of a few \citep{paschmann_energization_1980,tsurutani_upstream_1981}, a process that can be repeated until the particle can overcome the shock potential, which is often referred to as shock-surfing \citep[e.g.][]{shapiro_shock_2003}. Processes involving reflection preferentially inject ions since their Larmor radii are larger with respect to the shock width compared to that of the electrons, which instead get trapped in the shock transition region. Counter-streaming ions can induce microinstabilities, for example the modified two-stream instability \citep{mcbride_theory_1972,matsukiyo_modified_2003,matsukiyo_microinstabilities_2006} and the Whistler modes mentioned above, which help inject electrons and further ions. While PIC simulations have provided many important recent advances regarding injection, the detailed injection physics for shock acceleration of both electrons and ions is extremely complicated and depends on the detailed shock characteristics \cite[see][for a summary]{marcowith_microphysics_2016}.

\begin{figure*}
    \centering
    \includegraphics[width=1.0\linewidth]{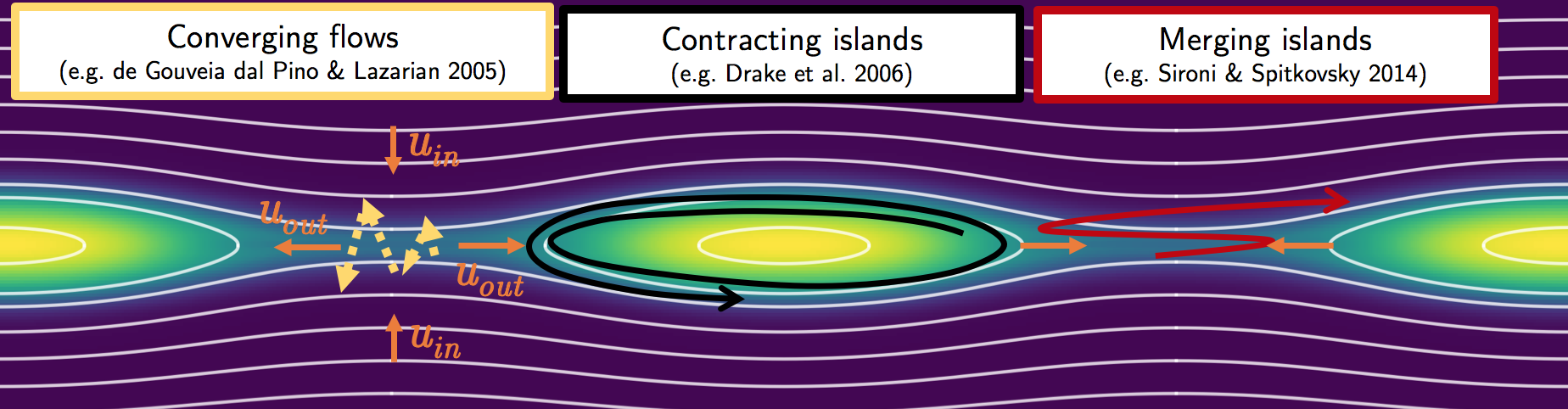}
    \caption{Schematic showing a site of magnetic reconnection. The background colourmap shows the plasma density (yellow denoting high density) from a simple resistive MHD simulation of a Harris current sheet, carried out using \textsc{pluto}\ \citep{mignone_pluto:_2007}, with magnetic field lines shown in white and the current density along the current sheet pointing into the page. The velocity vectors and yellow/black/red arrows showing crude particle trajectories are discussed in the models for particle acceleration at a reconnection site described in Section~\ref{sec:reconnection} and the labelled references. For actual particle trajectories produced from PIC simulations see e.g. fig. 2 of \citet{drake_electron_2006}, fig.~5~ of \citet{guo_particle_2015} and fig.~5 of \citet{sironi_relativistic_2014}.
    }
    \label{fig:reconnect}
\end{figure*}

\subsection{Magnetic Reconnection}
\label{sec:reconnection}
Magnetic reconnection has recently gained a lot of attention as a particle acceleration mechanism in jets, partly because it may help to explain how magnetic energy is converted into radiation in the bases of jets that are initially Poynting flux dominated  -- a phenomenon dubbed ``magnetoluminescence" by \cite{blandford_magnetoluminescence_2017}. The magnetic energy is liberated and converted into bulk motion, heat, and energetic particles.  A simple schematic of a reconnection site is shown in Fig.~\ref{fig:reconnect}, showing how converging field lines of opposite polarity approach each other and undergo resistive dissipation. The characteristic shape of the field lines means that the central reconnection region is often referred to as an `X-point' or `X-line'. Magnetic reconnection is often described using two main non-relativistic frameworks: Sweet-Parker reconnection \citep{parker_sweets_1957,sweet_neutral_1958}, where a long current sheet forms between the converging field lines, or Petschek reconnection \citep{petschek_magnetic_1964}, where slow shocks form just outside the X-point region. 

Sweet-Parker reconnection is extremely slow, even when kinetic effects are accounted for \citep{comisso_thermal-inertial_2014}, and it is well known that the solar flares in the sun require reconnection rates orders of magnitude faster than the Sweet-Parker value \citep[e.g.][]{isobe_reconnection_2009,kagan_relativistic_2015,galtier_introduction_2016}. Therefore, one of the fundamental questions relating to magnetic reconnection is how {\em fast reconnection} can take place, which is generally thought to happen via tearing/plasmoid instabilities that cause the current sheet to fragment into a series of magnetic islands \citep[e.g.][]{loureiro_instability_2007,bhattacharjee_fast_2009,uzdensky_fast_2010,huang_distribution_2012,loureiro_magnetic_2012,ni_fast_2015} and/or turbulent reconnection \citep[e.g.][]{lazarian_reconnection_1999,lazarian_astrophysical_2005,onofri_stochastic_2006,kowal_numerical_2009,loureiro_turbulent_2009,lazarian_turbulence_2012,kowal_statistics_2017,isliker_particle_2017}. Some of the studies discussed here focus on relativistic reconnection, defined as the regime where the magnetic energy density exceeds the rest mass energy density of the plasma, a regime that is relevant to jets. Fortunately, the non-relativistic models can be generalised to the relativistic case \citep{lyubarsky_relativistic_2005}. 

Particles can be accelerated in a number of different ways close to a reconnection site, as summarised by e.g. \cite{blandford_magnetoluminescence_2017,drake_computational_2018}. The first is by `direct' mechanisms in an electric field associated with the current sheet \citep{litvinenko_particle_1996,litvinenko_electron_1999,kirk_particle_2004,guo_efficient_2016}. Acceleration in this electric field is generally thought to be fairly inefficient \citep{dahlin_parallel_2016,dahlin_role_2017}, but it is probably an important injection mechanism \citep{lyutikov_particle_2016,li_particle_2016,comisso_particle_2018}. The second is by betatron acceleration, where the first adiabatic invariant ($p_\perp^2/B$) of a particle is conserved in a slowly-increasing magnetic field such that the particle energy increases, but this is generally unimportant in this case since $B$ is generally decreasing. The third is by {\em Fermi-type mechanisms} that can be first-order in $(u/c)$, where, in the case of relativistic reconnection, $u$ is a significant fraction of $c$. 

Magnetic reconnection creates a situation where two regions of opposing magnetic polarity approach each other at a velocity $u_\mathrm{in}$. This is a convergent flow that is closely analogous to that in a shock. If particles cross from the lower to upper regions they acquire an energy gain in much the same way as in DSA. The relative velocity in the rest frame of either of the converging two regions is $2 u_{\mathrm{in}}$, so if we average over an isotropic pitch angle distribution (factor of $1/3$) and consider the round trip (factor of 2) then we have $\langle \Delta E/E \rangle \sim 8/3 u_{\mathrm{in}}/c$, and $\beta = 1 + (8/3) (u_{\mathrm{in}} /c)$, identical to the case of Fermi II except that the energy change is first-order in $(u_{\mathrm{in}} /c)$. Using a simple approximation for particle escape, \cite{de_gouveia_dal_pino_production_2005,de_gouveia_dal_pino_particle_2015} derive $P=1-4u_{\mathrm{in}}/c$. Combining these values with equation~\ref{eq:energy_spec} gives a predicted spectrum of $s=2.5$, close to the inferred index from some synchrotron spectra. However, \cite{drury_first-order_2012} notes that these assumptions about escape do not really apply and reformulates the problem in terms of a compression ratio, $\rho_{\mathrm{out}}/\rho_{\mathrm{in}}$, where the subscripts denote densities in the outflowing and inflowing plasma. This leads to a spectral index that tends towards $s=1$ for the `maximally compressive' case, i.e. when $u_{\mathrm{in}} / u_{\mathrm{out}}$ is maximised. Reconnection is most compressive when it is efficient in converting magnetic energy to kinetic energy, which may be exactly what is needed to transition from a Poynting flux dominated jet to a kinetically dominated one. This is an important general point; a Fermi process in a reconnection site creates an intrinsic link between the release of magnetic energy and the acceleration of particles. In both these models, there is a similarity to shock acceleration in that there are reasonable grounds for expecting the particle escape probability to be constant with increasing energy.

Although these simple models are instructive, many of the advances in understanding Fermi-type processes close to reconnection sites have come through detailed PIC simulations. For example, \cite{drake_electron_2006} show that particles can undergo a Fermi process by following field lines around a contracting magnetic island, experience an energy boost of $\sim v_A/c$ each time the particle reaches the ends of the islands and reverses $x$ direction (see e.g. our Fig.~\ref{fig:reconnect}, Drake at al.'s fig. 2). \cite{sironi_relativistic_2014} have shown that the plasmoid instability fragments the current sheet into a series of magnetic islands of varying sizes; they find a variety of first-order Fermi-type mechanisms at work, with particles scattering between merging magnetic islands and the largest islands generally acting as reservoirs for the highest-energy particles \citep[see also][]{lyutikov_particle_2016,ball_electron_2018}. An important difference to Drury's picture is that the velocity in the energy gain is proportional to $u_{\mathrm{out}}$ which can be highly relativistic, whereas generally $u_{\mathrm{in}} \sim v_A \sim 0.1c$ \citep{lyubarsky_relativistic_2005,sironi_relativistic_2014}. \cite{guo_particle_2015} show using test particle MHD simulations that particles undergo a curvature drift along an electric field induced by the reconnection flows \citep[see also][]{Li_2017}. These Fermi mechanisms differ clearly from the DSA case, since the presence of self-generated turbulence is not necessarily needed to facilitate a first-order Fermi process.

Simulations have therefore been relatively successful in producing superthermal particles with power-law spectra from reconnection sites, but the energy regime probed is relatively low, up to around $100-1000$ times the thermal energy, and the maximum scale is a similar factor of the ion inertial length \citep[e.g.][]{Li_2017}. Ignoring losses, the maximum energy in a Fermi-type process in an individual island or X-line region will (optimistically) be set by the confinement condition, but if multiple magnetic islands exist then this also facilitates a stochastic process similar to Fermi II in which particles interact with multiple reconnection sites. This could be thought of as a `Fermi 1.5' process, since the particle gains energy in a first-order process until its gyroradius becomes comparable to the size of an individual reconnection site, before escaping and scattering off other islands. \cite{hoshino_stochastic_2012} argues that this process is first-order in $v_A/c$ since the particles preferentially interact with the outflowing plasma from the reconnection sites. \cite{comisso_particle_2018} have conducted simulations of a turbulent plasma in which particles are injected at X-points before undergoing second-order Fermi acceleration in the turbulent plasma; they find a spectral index of $s=2.9$ which is not universal and softer than in the more ordered situation considered by \cite{sironi_relativistic_2014}. Given that fast reconnection is likely to be inherently turbulent, it is important to consider the overall, 3D turbulent environment, since the details of the process clearly depend on the scales of the various reconnection sites/magnetic islands. PIC simulations and MHD simulations with test particles are therefore critical tools for understanding particle acceleration at reconnection sites. The particle spectral index from these simulations varies, but values for $s$ tend to lie in the range 1 to 2 for both $e^-e^+$ and $e^-$-ion plasmas \citep[e.g.][]{romanova_magnetic_1992,hoshino_suprathermal_2001,zenitani_generation_2001,larrabee_lepton_2003,lyubarsky_particle_2008,sironi_relativistic_2014,guo_particle_2015}.

\subsection{Other mechanisms}
The mechanisms we have described are the most commonly invoked to explain high-energy particles, but there are of course other ways in which particles can acquire highly superthermal energies. {\em Shear acceleration} is a form of Fermi acceleration - the flow is not converging, but the particle scatters across a shear layer such as that at the edge of a jet \citep[e.g.][]{rieger_shear_2004}. The particle distribution function can once again take the form of a power-law in momentum whose index is determined by how the time between scattering events depends on the momentum of the particle \citep{berezhko_kinetic_1981,berezhko_formation_1982}. The non-relativistic particle transport equation was derived by \cite{earl_1988}, and the problem has been since been revisited (in the case of relativistic flows) by \cite{webb_shear_2018,webb_shear_2019} and \cite{rieger_particle_2019}. Recent reviews are provided by \cite{rieger_introduction_2019,rieger_energetic_2019}. Shear acceleration may also be able to provide a `one-shot' energy boost by a factor of $\sim \Gamma^2$ to an existing population of high-energy particles encountered by the jet (\citealt{caprioli_espresso_2015}; see also \citealt{kimura_ultrahigh-energy_2018}). {\em Magnetic kink instabilities} can also produce nonthermal distributions of particles \citep[e.g.][]{alves_efficient_2018} and are interesting since the kink-mode operates in jets under certain conditions \citep{begelman_theory_1984,mignone_high-resolution_2010,tchekhovskoy_three-dimensional_2016,barniolduran2017}. 

\subsection{General Considerations}
Although the microscopic physics changes depending on which mechanism is in operation, the key underlying principles are similar. The dominant mechanism at a given point in a jet is set by the local physical conditions; we discuss this in Section~\ref{sec:discussion}. First, we make a few general comments. 

The first relates to the availability of different Fourier modes in magnetic field structure at the particle acceleration site. To accelerate a particle from, say, GeV to TeV energies requires structure in the magnetic field on the scale length of the respective Larmor radii - i.e. over the same dynamic range as the energy range in question. One great advantage of a shock in this regard is that it is a discontinuity, so, provided the shock is not too smoothed out, it is possible for every Fourier mode of turbulent magnetic field to be present at the shock, driven by the  self-regulated processes we described in Section~\ref{sec:nonlinearity}. It is of course possible that this condition is fulfilled by other mechanisms, but it depends critically on the spectrum of magnetic irregularities. In Kolmogorov turbulence, with $E(k)\propto k^{-5/3}$, where $k$ is the wavenumber of the turbulence and $E(k) dk$ is the energy density contained in the wavenumber range $(k,k+dk)$, the turbulence covers a wide range of scales, but one has to account for the fact that the energy contained at small scales (large $k$) is small compared to the energy density at the driving scale. 

Another related point concerns the spectral index of inferred particle energy distributions. We have implied throughout this section that DSA has an advantage over some of the other acceleration mechanisms not only in its first-order velocity dependence, but also in that the spectral index does not require fine-tuning and has a `universal' value. It is worth noting that a particle energy index of $s=2$ spreads the energy in superthermal particles out such that each decade in energy contributes an equal amount to the energy density. It is therefore possible that the implied value of $s=2-2.6$ inferred from synchrotron power-laws is a reflection of some self-regulating process that tries not to put the majority of energy in the high-energy particles, and may not provide deep physical insight into the specific mechanism. 

\section{Timescales and the maximum energy}\label{sec:max_energy}
The characteristic maximum energy achievable by a particle with charge $Ze$ being moved a distance $R$ through a $-\vec{u}\times\vec{B}$ electric field is set by the Hillas energy \citep{hillas_origin_1984}, given by $E_H = ZeuBR$,
which can be written as 
\begin{equation}
E_H = 10^{18}~\mathrm{eV}~\left(\frac{R}{1~\mathrm{kpc}}\right)~
\left(\frac{B}{1~\mu \mathrm{G}}\right)~
\left(\frac{u}{c}\right)~Z.
\label{eq:hillas}
\end{equation}
This can also be obtained by taking the time derivative of a magnetic flux $BR^2$, which gives a potential drop $uBR$ \citep[e.g.][]{lemoine2009,waxman_high_2011}. If the flow is relativistic, an additional factor of $\Gamma$ is included in estimates for the maximum energy \citep[e.g.][]{achterberg_particle_2001} depending on the details of the acceleration process, the geometry, the frame in which $B$ is defined, and the escape of the particle after acceleration. Equation~\ref{eq:hillas} becomes the confinement condition (Equation~\ref{eq:confinement}) from Section~3 if we set $u=c$, but the distinction is important. In DSA, $u$ is the shock velocity, whereas in magnetic reconnection it is some flow velocity comparable to $v_A$. Since $u<c$, it is, unsurprisingly, harder to accelerate a particle than to confine it and the accelerator must therefore be larger than the Larmor radius by a factor $c/u$.

\subsection{Acceleration timescale}
The Hillas criterion is necessary but not sufficient for acceleration to high energy. In reality the crucial aspect is a balance between the acceleration timescale and the dominant timescale of escape, adiabatic losses or radiative cooling. The general expression for the acceleration time is 
\begin{equation}
t_{\mathrm{acc}} \sim \frac{E}{\dot{E}_{\mathrm{acc}}} \sim \left < \frac{\Delta E}{E} \right > t_{\mathrm{cycle}},
\end{equation}
where $t_{\mathrm{cycle}}$ is either the collision time (for Fermi II) or crossing time (for DSA). A more specific timescale for DSA was derived in the SNR context by \cite{lagage_cosmic-ray_1983,lagage_maximum_1983} and \cite{drury_introduction_1983}, who find  
\begin{equation}
t_{\mathrm{acc,DSA}} = \frac{3}{u_s^2} \left( \frac{\chi D_u + \chi^2 D_d}{\chi - 1} \right),
\label{eq:tacc_dsa}
\end{equation}
where $D_u$ and $D_d$ are the upstream and downstream diffusion coefficients and $\chi$ is the shock compression ratio. As noted by \cite{blasi_maximum_2007}, the above formula is not appropriate for a CR-modified shock and the expression becomes more complicated, but it is suitable for our purposes. Except for in special situations \citep[e.g.][]{bell_flux_tube}, the Hillas energy is only reached in DSA at a parallel or oblique shock if Bohm diffusion applies. This can be demonstrated using equation~\ref{eq:tacc_dsa}. If we assume the upstream and downstream dwell times ($\propto D/u$) are identical, and set $t_{\mathrm{acc}}=R/u_s$, then for a strong, non-relativistic shock with $\chi=4$ we recover $E_{\mathrm{max}} = 3/8 (D_\mathrm{Bohm}/D_u) u B R$, where the exact preceding numerical factor is determined by the relationship between $D_u$ and $D_d$ as well as the value of the denominator in $D_\mathrm{Bohm}$. At a perpendicular shock, the particle can in principle drift along the shock for its entire length if the turbulence either side of the shock is just sufficient to counteract the $\vec{{\cal E}} \times \vec{B}$ drift, in which case the Hillas energy can be reached. However, in practice this is unlikely, so the particle must return to the shock multiple times to reach the Hillas limit, which is harder to achieve than for a parallel shock. Similar expressions to the non-relativistic case can be derived for the maximum energy at ultra-relativistic shocks \citep[e.g.][]{achterberg_particle_2001,bell_cosmic-ray_2018}, where $u_s\rightarrow c$ and $\chi\rightarrow 3$, but the generation of the turbulence becomes a severe limitation (see sections~\ref{sec:relativistic_shocks} and \ref{sec:growth})

The acceleration time for Fermi II actually has a very similar form to that for DSA; although the energy gain is smaller by a factor $(u/c)$, the collision time is also shorter than the shock crossing time and the acceleration time is given by 
\begin{equation}
t_{\mathrm{acc,F2}} \propto  \frac{D}{u_\mathrm{turb}^2}
\end{equation}
\citep[e.g.][]{jones1994}, where $u_\mathrm{turb}$ is the characteristic velocity of the turbulence/clouds/waves. Why, then, is Fermi II normally thought of as an inefficient process? The answer is that the characteristic velocity is generally lower than in DSA, so a larger containing volume by a factor $u_s/u_\mathrm{turb}$ is needed for a particle to reach the same energy. In supersonic jets, the turbulent velocity cannot be comparable to the bulk velocity, highlighting again why shocks and reconnection sites are  much more efficient at channeling magnetic or kinetic energy into superthermal particle distributions. Fermi II might however be important indisrupted, turbulent jets \citep[e.g.][]{manolakou_particle-acceleration_1999} or in the jet cocoon or lobe \citep[e.g.][]{begelman_theory_1984,hardcastle_high-energy_2009}.

The acceleration time for particle acceleration in reconnection sites depends on the details of the acceleration process. We might, however, expect similar arguments to DSA/Fermi II to apply depending on how ordered the process is. Detailed studies using test-particle simulations find $t_{\mathrm{acc,rec}} \propto (u_{\mathrm{rec}}/c)^{-\kappa}$ with $\kappa$ ranging from 2.1 to 2.4 \citep{del_valle_properties_2016}. This is close to the $u^{-2}$ dependence quoted above for DSA and Fermi II. 

\subsection{Instability Growth Times}
\label{sec:growth}
Acceleration timescales based on diffusion coefficients involve an implicit assumption that there is turbulent magnetic field on the scale of the Larmor radius of the particle. In the case of self-generated turbulence from streaming instabilities we must also consider the growth rate of the turbulence. For NRH/Bell turbulence, the process is self-regulating and the CRs drive turbulence on the scale of their respective Larmor radii which in turn scatters particles of that energy. The growth rate of the NRH instability is given by $\gamma_\mathrm{max} \sim 0.5 j_{cr} \sqrt{\rho/\mu_0}$ \citep{bell_turbulent_2004}, where $j_{cr} \approx n_{cr} e u_s$ is the magnitude of the CR current. The maximum particle energy is limited by the scale size and strength of magnetic field that can be generated. One way of thinking about this is that the scale size of the magnetic field is limited by the maximum displacement from the $-\boldsymbol{j} \times \boldsymbol{B}$ force associated with the CRs.  

In SNRs, this means that the maximum energy is limited by the time taken for magnetic field amplification and can be significantly below the Hillas limit \citep{zirakashvili2008,bell_cosmic-ray_2013}. In relativistic shocks, the general ability of CRs to generate turbulence is inhibited and the maximum energy is orders of magnitude lower than that expected from the Hillas energy \citep{bell_cosmic-ray_2018}. These difficulties might be bypassed in a few specific circumstances; for example, if there is pre-existing turbulence, if the geometry of the acceleration site is such that the maximum energy becomes independent of the scattering frequency \citep{eichler_energetic_1981}, or if there are multiple acceleration sites. Furthermore, this limit on the energy only applies if self-generated plasma instabilities are responsible for scattering the particles, so this is not applicable to certain scenarios associated with magnetic reconnection and Fermi II. However, quite generally, a particle will only be accelerated to an energy $E$ on a short timescale if there is structure in the magnetic field on the scale of the Larmor radius at that energy, regardless of the mechanism for generating this structure.

\subsection{Losses}
Radiative cooling limits the maximum energy if the radiative cooling timescale for the particles is short enough. The generic expression for any radiative loss timescale is 
\begin{equation}
t_\mathrm{cool, rad} = \frac{E}{\dot{E}_{\mathrm{rad}}},
\label{eq:rad_cool}
\end{equation}
where $\dot{E}_{\mathrm{rad}}$ describes the rate at which a particle radiates due to synchroton, inverse Compton or synchrotron self-Compton processes. The synchrotron power of an electron in a magnetic field with energy density $U_B=B^2/8\pi$ is 
\begin{equation}
\dot{E}_{\mathrm{sync}}= \frac{4}{3} \sigma_T c \gamma_e^2 \beta_e^2 U_B,
\end{equation}
where $\beta_e=v_e/c$ and $\sigma_T$ is the Thomson cross-section. 
Similarly, the inverse Compton cooling power is
\begin{equation}
\dot{E}_{\mathrm{IC}}= \frac{4}{3} \sigma_T c \gamma_e^2 \beta_e^2 U_{\mathrm{rad}},
\end{equation}
giving us the simple result that the maximum energy loss rate is just determined by the energy density of the dominant field interacting with the electron. The value of $U_{\mathrm{rad}}$ is often uncertain and will vary with height in the jet. For example, close to an accretion disc the disc luminosity may be dominant, but further from the jet base the electrons' own synchrotron radiation or the CMB dominates. Similarly, we might expect $U_B$ to vary dramatically along a jet if it transitions from Poynting flux dominated to kinetically dominated. 

\begin{figure*}
    \centering
    \includegraphics[width=0.9\linewidth]{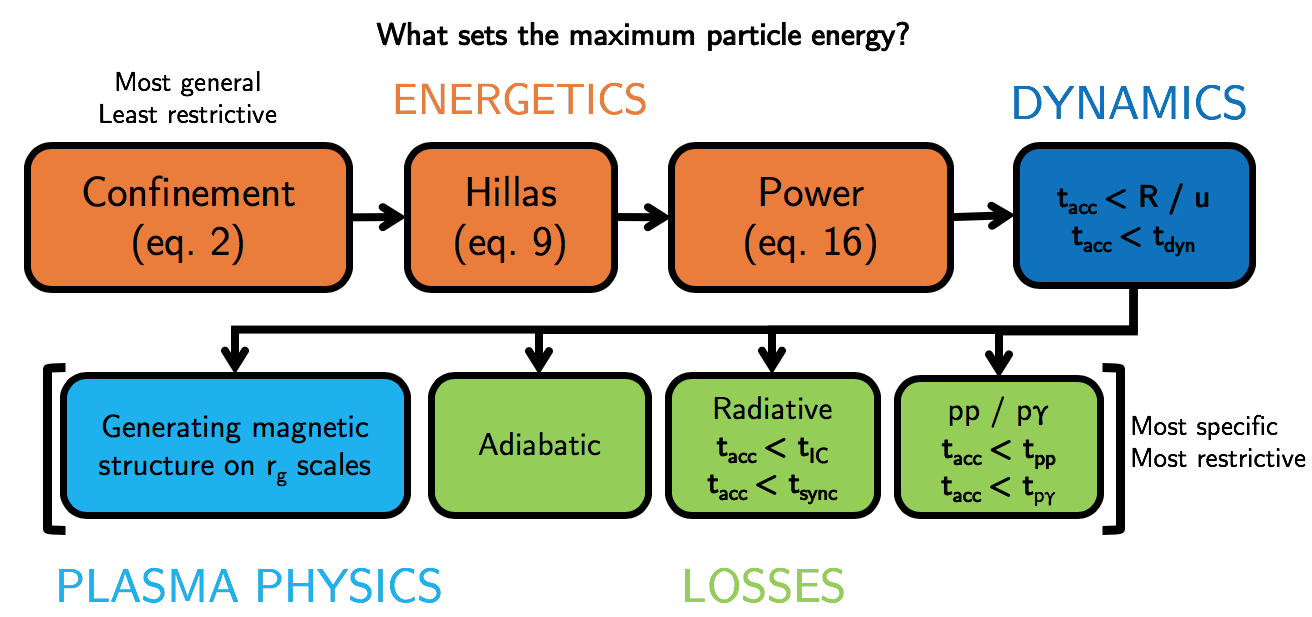}
    \caption{A maximum energy `ladder'. The flowchart starts at the most general, least restrictive condition for the maximum energy and gradually incorporates more detailed physics. The $pp$ and $p\gamma$ channels are only for protons whereas the adiabatic and radiative channels apply to protons, ions and electrons.  The limiting factor for the maximum energy depends on the acceleration process, the energy densities $U_B$ and  $U_{\mathrm{rad}}$, the adiabatic loss timescale, the density and the detailed plasma physics that leads to scattering of particles.}
    \label{fig:max-energy-ladder}
\end{figure*}

Protons and ions also cool. The proton synchroton and inverse Compton cooling times can be derived by considering the mass ratio $m_p/m_e \approx 1837$, which, for a given energy, leads to a  cooling time a factor $(m_p/m_e)^4 \approx 10^{13}$ longer than for electrons \citep[e.g.][]{begelman_consequences_1990,aharonian_tev_2000}. Thus, proton synchrotron and inverse Compton radiation are often negligible, although may be important in strong $\gtrsim100$G magnetic fields in the bases of AGN jets \citep{aharonian_tev_2000,aharonian_proton-synchrotron_2002,biteau_progress_2020} or in GRBs \citep{waxman_high-energy_2001,gupta2007}. Nonthermal protons also lose energy via the same $pp$ and $p\gamma$ channels that produce gamma-ray emission. The cross-section for $p\gamma$ processes depends strongly on the particle energy and shape of the radiation field; for low temperature photon fields such as the CMB, radiation losses are only significant at ultrahigh energies \citep[e.g.][]{alves_batista_effects_2015}, whereas close to an accretion disc the strong UV radiation field can cause fast $p\gamma$ losses at $\sim$PeV energies \citep[][see their fig.~2]{begelman_consequences_1990}. Additionally, ions can photodisintegrate leading to a reduction in energy \citep{stecker_effect_1968,stecker_photodisintegration_1999}. Inelastic collisions with thermal protons also result in energy losses, but the associated timescale is quite long: $t_{pp} \approx (\tfrac{1}{2} \sigma_p n_p c)^{-1} \sim 600$~Gyr for a density of $n_p=10^{-4}$~cm$^{-3}$ \citep{sikora_electron_1987}, where $\sigma_p\approx4\times10^{-26}$~cm$^2$ is the cross-section. Losses from $pp$ collisions can nevertheless be the dominant proton cooling mechanism for $\lesssim 10$~TeV energies. For protons, the regimes in which the different types of losses are expected to dominate are summarised by \cite{begelman_consequences_1990}, showing a strong dependence on $U_\mathrm{rad}$, $n_p$, $U_B$ and particle energy.

\subsection{Power requirement}
The Hillas energy can be used to derive a maximum energy based on the power of a source \citep{lovelace_dynamo_1976,waxman_cosmological_1995,blandford_acceleration_2000,lemoine_electromagnetic_2010}. This power requirement is sometimes referred to as a Hillas-Waxman-Lovelace limit. It can be derived by considering the `magnetic power' passing through a surface of cross-sectional area $R^2$, which is $Q_B = (B^2/8\pi)~R^2~u$, where $u$ is the bulk velocity through the surface. Combining this with equation~\ref{eq:hillas} then gives
\begin{equation}
Q_{B,\mathrm{min}} = 10^{44}~\mathrm{erg~s}^{-1} 
\left( \frac{E}{10 \mathrm{EeV}} \right)^2
\left( \frac{u}{0.1c} \right)^ {-1}
~Z^{-2}.
\label{eq:power}
\end{equation}
This limit is due to the maximum electric field produced by a source with a certain magnetic power, assuming that we arrange for optimal acceleration conditions. A less restrictive power requirement is sometimes derived by setting $u=c$, but equation~\ref{eq:power} applies quite generally irrespective of acceleration mechanism. For example, following e.g. \cite{potter_using_2017}, in a reconnection site $u=u_{in}$ and $R^2$ is the surface area of the current sheet, whereas in shock acceleration $u=u_s$ and $R$ is roughly the shock size. We can convert this minimum magnetic power into a kinetic power using a parameter $\eta$ such that $Q_{k,\mathrm{min}} = \eta^{-1} Q_{B,\mathrm{min}}$, where $\eta$ represents the partitioning between kinetic and magnetic energy densities; in shock acceleration, $\eta$ can be thought of as an efficiency of magnetic field amplification at the shock. 

For most of the observations described in Section~2, the maximum particle energy is significantly lower than that associated with equation~\ref{eq:power}, with the important consequence that either the acceleration process or loss processes are limiting the maximum energy. However, for UHECRs, where the particle acceleration process is presumably stretched to its limits, this limit becomes important and already rules out a large number of different sources such as starburst winds, AGN disc winds or microquasar jets. In fact, jets in AGN and GRBs are two of the few known sources that, for reasonable efficiency parameters, satisfy $Q_k>Q_{k,\mathrm{min}}$ for 10 EeV particles.

\subsection{A maximum energy ladder} 
Having explored what factor limits the maximum particle energy we are able to build a `maximum energy ladder': a set of gradually more restrictive criteria that determine the maximum energy for a given mechanism for a specific set of physical conditions. An example diagram is shown in Fig.~\ref{fig:max-energy-ladder}. The first four criteria are based purely on the size, speed and energetics of the accelerator and at this point one can remain agnostic about the detailed acceleration process. The bottom half of the flowchart refers to more involved physics, and in this case the limiting factor will depend on the physical parameters in the plasma, for example $t_{\mathrm{acc}}$, $\vec{B}$ and $U_{\mathrm{rad}}$. The diagram illustrates why it is crucial not only to have reasonable constraints on the physical conditions in a given source, but also a good understanding of the detailed plasma instability growth rates (if applicable) and acceleration timescales for the most likely acceleration mechanism.

\section{Cosmic Rays and Neutrinos}
\label{sec:multimessenger}

\begin{figure*}
    \centering
    \includegraphics[width=0.85\linewidth]{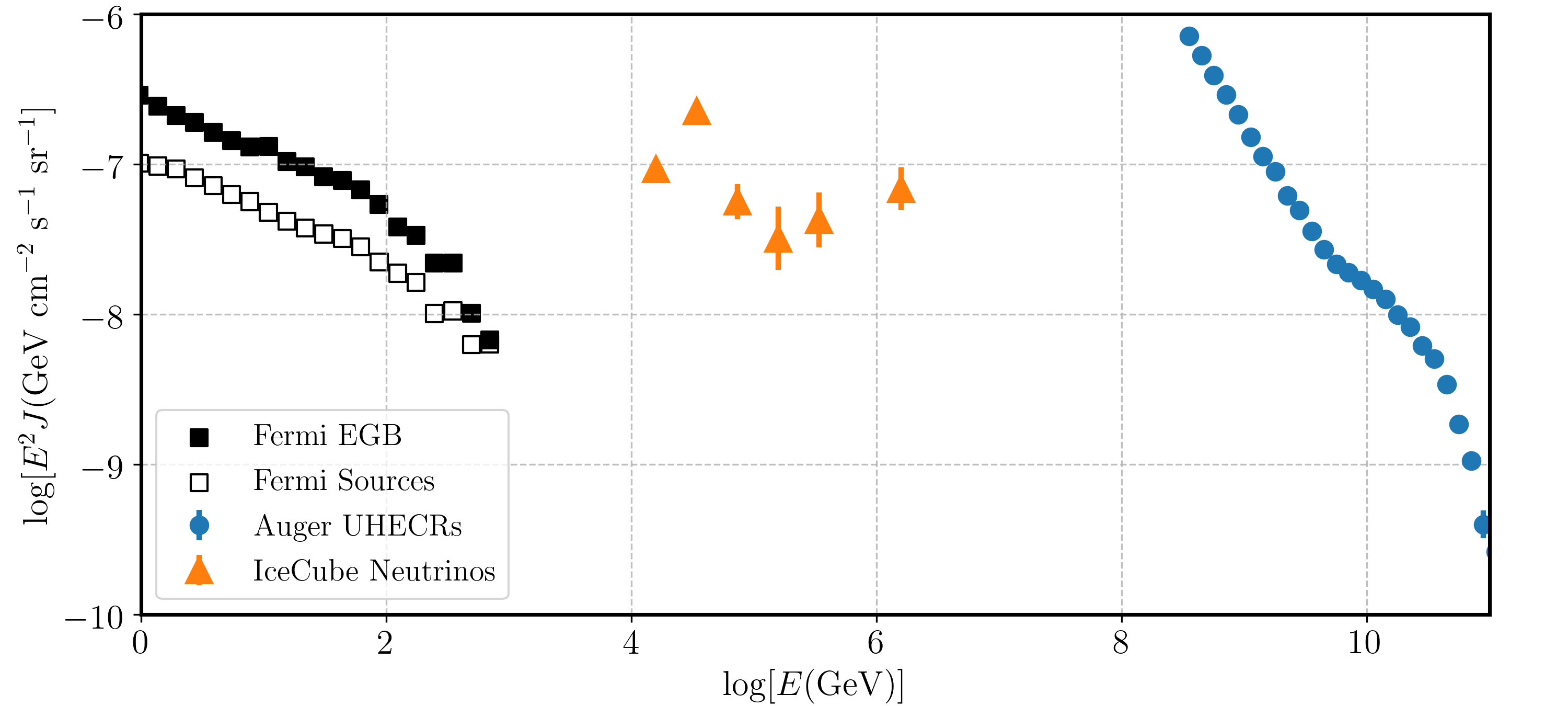}
    \caption{Spectra showing multimessenger signals of of high-energy particle acceleration (L to R: photons, neutrinos and hadrons) arriving at Earth. References for the data sources are given in Section~\ref{sec:multimessenger}. The {\sl Fermi} data is split into the total Extragalactic Gamma-ray Background (EGB) and the contribution from individually resolved sources, as described by \cite{ackermann_spectrum_2015}. The plot is inspired by similar figures from \cite{fang_linking_2018} and \cite{ahlers_neutrino_2018}. Spectra are plotted in units of $E^2~J$ so that a flat line contains equal total energy in each decade of particle energy. Astrophysical jets at least contribute to the flux and may even produce all three signals.}
    \label{fig:multi-spec}
\end{figure*}

In Fig.~\ref{fig:multi-spec}, we show differential particle fluxes ($J$) arriving at Earth, multiplied by $E^2$, for three types of particles: gamma-ray photons, neutrinos and hadrons (CRs). The figure is based on similar plots produced by, e.g.,  \cite{fang_linking_2018} and \cite{ahlers_neutrino_2018}. The data are from the {\sl Fermi} Large Area Telescope \citep{ackermann_spectrum_2015}, the IceCube Neutrino Observatory\citep{aartsen_combined_2015} and the Pierre Auger Observatory (PAO) \citep{pierre_auger_collaboration_combined_2017}, respectively, and the units are chosen so that a flat line represents equal amounts of energy per logarithmic energy bin. The contribitions of UHECRs (at $>$EeV energies),  high-energy IceCube neutrinos (at TeV to PeV energies) and the extragalactic gamma-ray background (at GeV to TeV energies) are comparable. This behaviour might initially appear to suggest universal physics; however, at least in the UHECR case the power-law is quite steep and so will inevitably cross a given value of $E^2J$. However, as has been pointed out by numerous authors \citep[e.g.][]{mannheim_high-energy_1995,waxman_high_1997,bhattacharjee_origin_2000,fang_linking_2018,ahlers_neutrino_2018,alves_batista_cosmogenic_2019} the three sets of high-energy particles may have common origins and extragalactic jets are plausible acceleration sites for protons, nuclei and their decay products.

\subsection{Neutrinos}
High-energy protons produce pions via interactions with photons or proton-proton collisions, which then decay to produce gamma-ray photons (in the case of $\pi^0$) or leptons and neutrinos  (in the case of $\pi^{\pm}$). Neutrinos are therefore sometimes considered to be a ``smoking gun'' of hadronic acceleration, making them useful messengers for studying jet composition and connections to UHECRs. The  {\sl IceCube} neutrino detector has proved to be a game-changer in the field of high-energy neutrino astrophysics. {\sl IceCube} has detected PeV neutrinos \citep{aartsen_first_2013,aartsen_observation_2014,collaboration_evidence_2013}, for which jets from GRBs or AGN are possible sources. Tentative coincidences have been investigated between PeV neutrino events and AGN sources \citep{kadler_coincidence_2016,fraija_study_2018}. Jet sources may also produce the diffuse TeV to PeV flux observed by {\sl IceCube} and shown in Fig.~\ref{fig:multi-spec}, although the maximum contribution to this diffuse flux from AGN in the 2nd {\sl Fermi-LAT} AGN catalog is $27\%$ \citep{aartsen_contribution_2017}.

Recently, {\sl IceCube} detected neutrinos spatially coincident with the BL Lac-type blazar TXS 0506$+$056 as it was undergoing a gamma-ray flare \citep{icecube_2018a,icecube_2018b}. They infer a neutrino energy of $\sim300$~TeV and maximum proton energies in the range $10^{14}-10^{18}$~eV. While this is some way off explaining the highest energy CRs, the {\sl IceCube} result is one of the clearest signals yet that jets accelerate hadrons to very high energies, and demonstrates the potential of multimessenger observations for probing conditions in extragalactic jets. However, this detection does not necessitate that the hadrons are significant in terms of the overall jet energy budget or that the gamma-rays from blazars are hadronic in origin. 

\subsection{UHECRs}
CRs were first detected over 100 years ago \citep{hess_1912}, even before astrophysical jets, but it is more recent progress that is relevant to this review. The CR spectrum extends smoothly across many decades in energy and takes the form of a power-law with various breaks \citep[e.g.][]{hillas_cosmic-ray_2006}. The CR power-law provides one of the motivations for seeking a power-law distribution for $n(E)$ in section~\ref{sec:theory}. Galactic CRs, with energies up to and possibly beyond the `knee' in the CR spectrum at PeV energies, are thought to be accelerated by SNRs \citep[e.g.][]{bell_particle_2014}, although microquasars may also contribute \citep{heinz_cosmic_2002,fender_energization_2005,cooper_cosmic_2020}. However, the origin of ultrahigh energy CRs (UHECRs), with energies extending beyond $10^{20}$~eV, is still unknown. Jets have been suggested as possible acceleration sites, both in various classes of AGN \citep{hillas_origin_1984,norman_origin_1995,fang_linking_2018} as well as GRBs  \citep{waxman_cosmological_1995,waxman_high-energy_2001}. Deflections by  intervening magnetic fields \citep[e.g.][]{sigl_ultrahigh_2003,taylor_uhecr_2011,farrar_deflections_2017} and the attenuating effect of photopion, pair-production and photodisintegration interactions \citep[e.g.][]{greisen_end_1966,zatsepin_upper_1966,stecker_photodisintegration_1999,alves_batista_crpropa_2016} make robust directional associations difficult. Simple energetic arguments, such as those discussed in the previous Section, favour extreme sources \citep{hillas_origin_1984,waxman_cosmological_1995,blandford_acceleration_2000}, and thus systems with jets remain natural candidates due to their enormous kinetic powers. 

\begin{figure*}
    \begin{subfigure}{1\textwidth}
          \centering
          \includegraphics[width=\linewidth, clip=true, trim=1in 1in 1.5in 1in ]{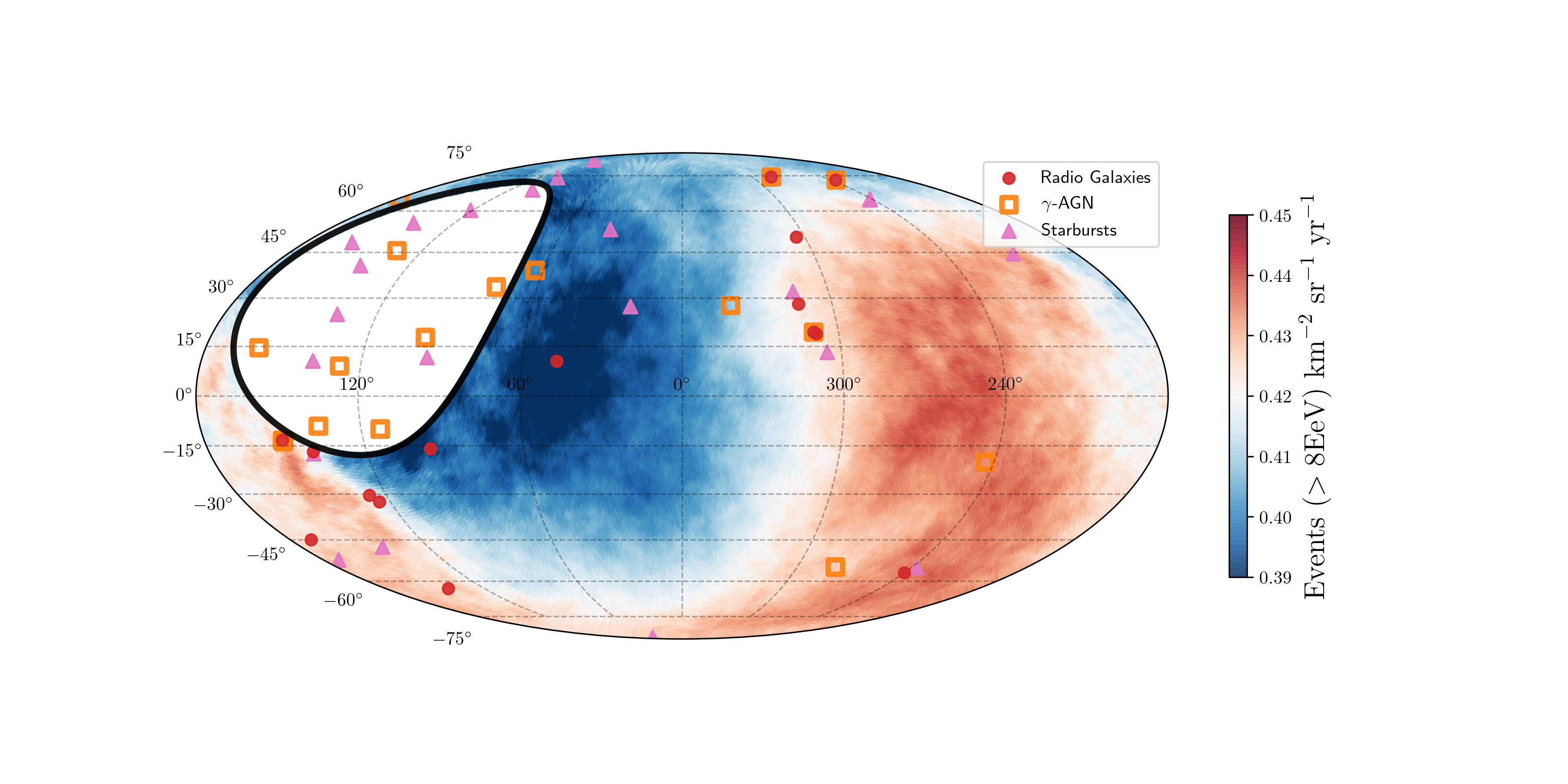}
    \end{subfigure}
    \begin{subfigure}{1\textwidth}
          \centering
          \includegraphics[width=\linewidth, clip=true, trim=1in 1in 1.5in 1in]{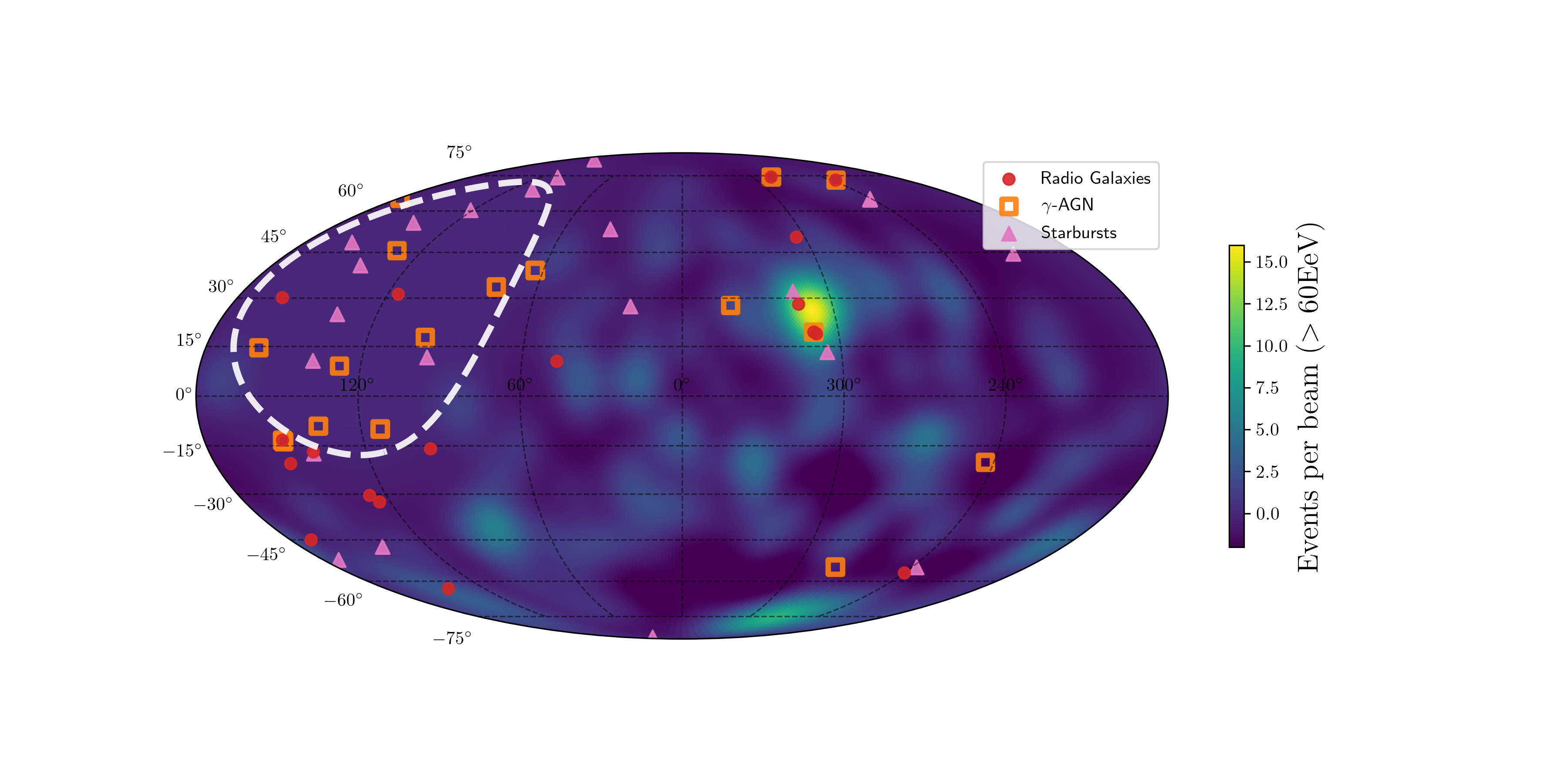}
    \end{subfigure}
    \centering
    \caption{{\sl Top:} Mollweide projection of UHECR fluxes above 8EeV in Galactic coordinates from PAO \citep{pierre_auger_collaboration_observation_2017}. A dipole anisotropy is observed in the data. {\sl Bottom:} Mollweide projection of the anisotropic excess events per beam above 60~EeV in Galactic coordinates from PAO \citep{pierre_auger_collaboration_indication_2018}. In both plots, the PAO exclusion zone in the northern hemisphere is marked. We also plot the gamma-ray AGN and SBG samples from \citep{pierre_auger_collaboration_indication_2018} and luminous ($\nu L_\nu > 2\times10^{40}$erg~s$^{-1}$) radio galaxies within 100~Mpc from the \cite{van_velzen_radio_2012} catalogue.}
    \label{fig:uhecr-map}
\end{figure*}

Recent advances have come from the combined power of the UHECR observatories, in particular the Telescope Array (TA) and the Pierre Auger Observatory (PAO). TA reported the detection of a `hotspot' in the Northern sky with a spread of $\sim20^{\circ}$ \citep{abbasi_indications_2014} while PAO have found a clear dipole above 8 EeV \citep[][hereafter PAO17]{pierre_auger_collaboration_observation_2017} and more tentative indications of anisotropy above $39$EeV and $60$EeV \citep[][hereafter PAO18]{pierre_auger_collaboration_indication_2018}. Fig.~\ref{fig:uhecr-map} shows Mollweide projections of the data from PAO17 and PAO18, in Galactic coordinates, with positions of some radio galaxies, blazars and starburst galaxies marked. PAO18 found that the anisotropic component could be associated with extragalactic gamma-ray sources, reporting correlations with starburst galaxies and AGN. Starburst galaxy `superwinds' have been proposed as UHECR sources \citep[e.g.][]{anchordoqui_acceleration_2018}, but these winds are generally too slow and low-power to meet the power requirement \citep[equation~\ref{eq:power}, see also][]{romero_particle_2018,matthews_fornax_2018}. However, starburst galaxies are expected to host large numbers of GRBs \citep{chary_are_2002,christensen_uv_2004,floch_probing_2006,becker_cosmic_2009}. GRBs generally offer better prospects for UHECR acceleration than starburst winds, although the required efficiencies are high \citep{waxman_cosmological_1995,waxman_high_2011,eichler2011,lemoine_ultra-high_2018,alves_batista_review_2019}.

An alternative is that UHECRs are accelerated by AGN jets. Radio galaxies have long been suggested as UHECR sources and recent work suggests they may be able to explain the data from PAO; for example, \cite{eichmann_ultra-high-energy_2018} suggest a combination of Cygnus A and Centaurus A, while \cite{matthews_fornax_2018} propose Fornax A and Centaurus A. One of the problems with accelerating UHECRs in AGN and GRB jets is that both internal and termination shocks are relativistic and this poses severe problems for acceleration to EeV energies \citep{lemoine_electromagnetic_2010,reville_maximum_2014,bell_cosmic-ray_2018}. One possibility in AGN is that non-relativistic shocks form in the backflows in the jet lobe/cocoon, offering more conducive conditions for UHECR acceleration \citep{matthews_2019}. Other proposed mechanisms for UHECR acceleration in AGN include magnetic reconnection in blazars \citep{giannios_uhecrs_2010} and Fermi II acceleration in the lobes of radio galaxies \citep{hardcastle_high-energy_2009,hardcastle_which_2010}. 

UHECRs are interesting for our purposes because they stretch the theory of particle acceleration to its limits. However, it is extremely challenging to link the physics of UHECR acceleration to that of, e.g., GeV synchrotron electrons due to the different factors limiting the maximum energy and the vast range of Larmor radii involved. Progress is likely to come from multimessenger approaches that combine neutrino, gamma-ray and hadronic signals as well as the combined power of future TA and PAO datasets. Further constraints on UHECR composition and the intervening magnetic field will also prove crucial for identifying UHECR sources. 

\section{Discussion: A Journey Along A Jet}
\label{sec:discussion}

\begin{figure*}
    \centering
     \includegraphics[width=0.9\linewidth]{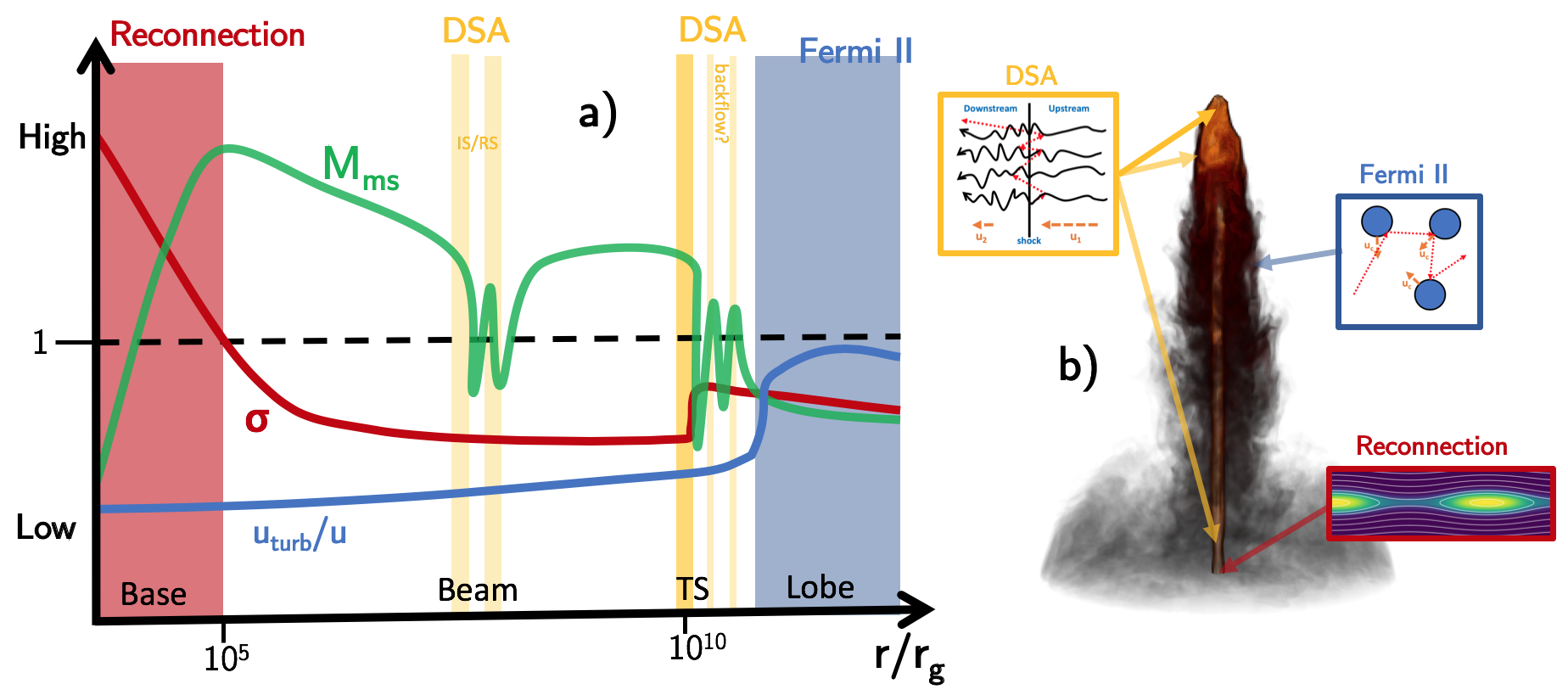}
    \caption{a) Schematic showing a possible profile along an AGN jet of three few key physical quantities defined in section~\ref{sec:discussion}: magnetisation, $\sigma$, magnetosonic Mach number, $M_{ms}$, and  the ratio of the bulk and turbulent velocities. Regions where certain mechanisms are expected to dominate are shaded. TS denotes termination shock. The figure is intended as a guide to aid the discussion in Section~\ref{sec:discussion}. The sketched profiles are informed by a combination of observational constraints and numerical simulations \citep[see][for examples of more detailed profiles]{potter_using_2017,matthews_2019,chatterjee_accelerating_2019}. IS/RS denotes internal shocks/reconfinement shocks. b) shows a volume rendering of the jet tracer, a passive scalar that tracks jet material, from a relativistic hydrodynamic simulation of an AGN jet, carried out using \textsc{pluto}. The rendering is shown for illustrative purposes and some possible mechanisms and sites for particle acceleration are labelled.
    }
    \label{fig:jet-profile}
\end{figure*}

Let us now consider the journey made by jet plasma, from a launching point near the compact object, through, e.g., a termination shock and into a lobe or cocoon. The plasma can be described at any point along this journey by a few key variables: the bulk Lorentz factor, $\Gamma$, the sonic and \alfven ic Mach numbers, the plasma-$\beta$, the magnetic field $\vec{B}$ and the velocity field $\vec{u}$. However, for the purposes of this discussion, we focus particularly on three key parameters that are essentially ratios of the various energy densities: 
\begin{itemize}
    \item $M_{ms}$, the magnetosonic Mach number of the flow;
    \item $\sigma=B^2/(4 \pi \Gamma \rho c^2)$, the magnetisation parameter, or the ratio of Poynting flux to mass-energy flux (in the observer frame);
    \item $u_\mathrm{turb}/u$, the ratio of the turbulent and bulk velocities.
\end{itemize}
The magnetosonic Mach number is defined as $M_{ms}= (\Gamma u) / (\Gamma_{ms} c_{ms})$ where $c_{ms}=\sqrt{v_A^2+c_s^2(1-v_A^2/c^2)}$ is the magnetosonic speed in a relativistic flow \citep{cohen_studies_2014}, and $\Gamma_{ms}$ is the Lorentz factor associated with $c_{ms}$. When the flow is magnetically dominated, $c_{ms} \approx v_A$, whereas at low magnetisations $c_{ms} \approx c_s$. We use $M_{ms}$ because it takes into account both the magnetic and thermal energy densities and governs the nature of the shocks that form \citep{marti_internal_2016}. These three parameters determine which particle acceleration processes are likely to operate most efficiently. A schematic showing how they might vary along a jet length and which processes are expected to dominate in each regime is shown in Fig.~\ref{fig:jet-profile}. 

We assume a powerful jet similar to those associated with FRII radio galaxies and powerful blazars that is also applicable to other situations where a persistent jet is produced. We assume that the jet is initially highly magnetised, as expected from simulations of jet launch \citep{mckinney_general_2006,komissarov_magnetic_2009} and constraints from radiative losses and variability in blazars \citep{potter_using_2017,morris_feasibility_2018}. The magnetisation parameter can change dramatically along a jet if it undergoes a transition from magnetically dominated near the base to kinetically dominated on large scales. Studies have suggested that this transition to $\sigma \lesssim 1$ occurs at around $10^5~r_g$ ($\approx0.5$pc for a $10^8M_\odot$ black hole) in blazar jets \citep{potter_using_2017,chatterjee_accelerating_2019}. In the magnetically dominated regime, reconnection is likely to be important. Fast, relativistic reconnection is a good candidate mechanism for explaining a few key observational constraints from Section~\ref{sec:obs}, that is: the fast variability observed from compact jet regions, the apparent need to dissipate magnetic energy in these regions and the requirement for {\sl in situ} particle acceleration along the jet length. The region of parameter space where reconnection is most likely to accelerate particles efficiently is shaded red in Fig.~\ref{fig:jet-profile}. 

Not all jets start off highly magnetised, in particular those in protostellar systems and some microquasars, while even the Doppler-boosted emission from blazars may originate outside the high-$\sigma$ region \citep[e.g.][]{sikora_are_2005}. Thus, the regime where $\sigma<1$ in Fig.~\ref{fig:jet-profile} can either correspond to regions further along the jet than the magnetised base, or to jets that are not initially Poynting flux dominated. In this lower magnetisation regime, acceleration at reconfinement shocks and internal shocks becomes important. We define internal shocks as those that are roughly planar and created by time-varying jet activity in which shells of different density and velocity in the jet collide with one another \citep[e.g.][]{rees_unsteady_1994,spada_internal_2001}. Reconfinement shocks are caused by the confining pressure of the cocoon/ambient medium, which drives an oblique shock into the jet  \citep[e.g.][]{falle_self-similar_1991,komissarov_evolution_2003}, and may disrupt the jet \citep{gourgouliatos_reconfinement_2018}. Both of these classes of shock appear to accelerate superthermal particles. Internal shock models are invoked in GRBs  \citep{rees_unsteady_1994,kobayashi_grb_1997,piran_physics_2004}, microquasars \citep{jamil_ishocks:_2010,malzac_internal_2013,drappeau_internal_2015} and AGN jets  \citep{spada_internal_2001,ghisellini_low_2002,bai_radio/x-ray_2003} to explain some of the observed temporal behaviour. Reconfinement shocks may explain the appearance of knots and quasi-periodic brightenings along the jet length in radio galaxies \citep{stawarz_dynamics_2006,nalewajko_dissipation_2012,hardcastle_deep_2016,levinson_reconfinement_2017}. Knots and spatially intermittent emission may also be explained by shocks from the jet interacting with stellar winds \citep{wykes_internal_2015}, a process that may also produce gamma-ray emission \citep{araudo_gamma-ray_2013}.

Jets can be disrupted and decelerated by entrainment and mass-loading \citep{komissarov_theoretical_1990,bowman_deceleration_1996,laing_dynamical_2002,perucho_numerical_2007} or instabilities \citep{hardee_stability_1995,tchekhovskoy_three-dimensional_2016,gourgouliatos_reconfinement_2018}. A low power or highly mass-loaded jet resembles a subsonic or transonic plume of radiating material rather than a focused supersonic beam. However, if the jet avoids disruption and remains at least moderately supersonic, the majority of its remaining kinetic power is dissipated through a strong shock. In the case of steady supersonic jets, the jet terminates in a reverse ``termination" shock responsible for the bright hotspots observed in radio galaxies. In transient events, the jet instead sweeps up surrounding material into a relativistic fireball \citep{blandford_fluid_1976,rees_relativistic_1992,waxman_high-energy_2001} and the particle acceleration occurs at a forward shock thought to be responsible for the afterglows seen in GRBs \citep{waxman_cosmological_1995,yost_study_2003,van_eerten_gamma-ray_2013}, although emission from reverse shocks can be observed at early times in long-duration GRBs \citep{laskar_reverse_2016,laskar_vla_2018,bright_detailed_2019}. Both forward and termination shocks will generally produce efficient  particle acceleration as the shock velocity can be large, so the acceleration time is short, and a large fraction of the jet energy can be transferred to superthermal particles. The termination shock is labelled in our schematic in Fig.~\ref{fig:jet-profile}, and coincides with a sharp drop in $M_{ms}$, as well as an increase in magnetic field due to amplification and compression at the shock. 

Inside a supersonic jet, $u_{\mathrm{turb}}/u$ must be small, else the turbulence would also be supersonic and would rapidly heat the plasma and decrease the Mach number. Beyond the termination shock, or in the plumes of disrupted jets, the flow becomes subsonic or transonic, and less ordered. Thus, in jet lobes and cocoons $u_{\mathrm{turb}}/u$ might feasibly be close to unity. However, fast bulk flows can still occur, a prominent example being the backflows of radio galaxies, which can produce shocks and accelerate particles \citep{saxton_complex_2002,matthews_2019}. The shocks in backflows may be ideal candidates for accelerating UHECRs \citep{matthews_proc2019,matthews_2019}. More generally, however, the cocoons and lobes around jets are probably fairly slow accelerators of superthermal particles since the characteristic values of $u$, $u_{\mathrm{turb}}$ and $v_A$ are low. The low value of the velocities in this regime is in contrast to a) the velocity of the jet, which is the characteristic velocity associated with DSA at the termination shock, and b) the \alfven\ speed in a Poynting flux dominated jet, which is the characteristic velocity associated with relativistic reconnection near the jet base.

\section{Concluding Remarks}
\label{sec:conclusions}
We have described how particles can be lifted out of the thermal pool and accelerated to high energy, usually via Fermi-type processes that allow them to move stochastically through a $-\vec{u} \times \vec{B}$ electric field. We have shown that there are a number of mechanisms, of which shock acceleration and reconnection are the most notable, that produce power-law momentum distributions of superthermal particles. We have discussed what limits the maximum energy of protons and electrons and explored how jets may produce multimessenger signals of particle acceleration. 

The overall future outlook for understanding particle acceleration in jets is bright; a wealth of observational data is becoming available and these data are truly `multimessenger'. In the last two years alone, there have been robust detections of particle acceleration in the afterglow from a gravitational wave source \citep{abbott_gravitational_2017} and more tentative associations between jetted sources and the production of UHECRs \citep{pierre_auger_collaboration_indication_2018} and neutrinos \citep{icecube_2018a,icecube_2018b}. The Event Horizon Telescope has recently obtained stunning images of the horizon-scale environment around the central black hole in M87 \citep{eht_first_2019}, an AGN that possesses one of the best studied, and first discovered \citep{curtis1918} astrophysical jets. Longer term, the Cherenkov Telescope Array will offer unprecedented sensitivity in gamma-rays from approximately 0.1 to 100 TeV. On the computational side, recent advances have allowed PIC simulations to be run in 3D with impressive results \citep{riquelme_electron_2011,sironi_particle_2011,sironi_relativistic_2014,guo_particle_2015,crumley_kinetic_2019} , while (GRM)HD simulations of relativistic jets provide insights into the magnetisation, morphology and stability of the jets and lobes on a range of scales \citep[e.g.][]{krause_very_2005,mignone_high-resolution_2010,van_eerten_gamma-ray_2012,van_eerten_gamma-ray_2013,mendygral_mhd_2012,hardcastle_numerical_2013,hardcastle_numerical_2014,yoon_global_2015,yoon_formation_2016,english_numerical_2016,tchekhovskoy_three-dimensional_2016,moscibrodzka_general_2016,barniolduran2017,liska_large-scale_2018,chatterjee_accelerating_2019}. 

Despite all this exciting progress, questions remain plentiful: Are jets leptonic, or lepto-hadronic? How important is entrainment? Why are many jets apparently `magnetoluminescent' at their bases? What limits the maximum energy in different sources? Where do UHECRs and PeV neutrinos come from? What creates the high-energy hump in blazars? Which particle acceleration mechanisms matter most in GRBs? What is the microphysics of particle acceleration at shocks and reconnection sites? How is TeV gamma-ray emission produced in microquasars? What mechanism produces the rapid variability and flaring in blazars? In many cases where nonthermal emission is observed, the origins of the nonthermal particles is not fully known and it is difficult to distinguish between the possible acceleration processes.

A combination of theoretical and observational advances will be needed to fully understand the particle acceleration process in jets. Other areas of astrophysics, involving the Sun, solar wind, supernova remnants and pulsar wind nebulae provide important comparison points for the jets community. For example, we can study magnetic reconnection in the Sun \citep{priest_magnetic_2002,dalla_particle_2005,zharkova_recent_2011,mckenzie_observational_2011} and striped wind of pulsars \citep{lyubarsky_reconnection_2001,petri_magnetic_2007,cerutti_extreme_2012}, or DSA in the bright rims of supernova remnants \citep{long_chandra_2003,vink_magnetic_2003,cassam-chenai_blast_2007,uchiyama_extremely_2007,bell_cosmic-ray_2013} and in the solar system \citep{tsurutani_upstream_1981,eichler_energetic_1981,wenzel_characteristics_1985,ellison_diffusive_1987,burgess_particle_2007,burgess_ion_2012}. Each of these environments provides a laboratory that can be used to learn about the same particle acceleration mechanisms thought to operate in astrophysical jets. 

\section*{Acknowledgements}
We would like to thank R. Fender, R. Wijers, A. Araudo, M. Hardcastle and J. Bright for helpful discussions. JM acknowledges a Herchel Smith Research Fellowship at Cambridge. This work was also supported by the Science and Technology Facilities Council under consolidated grant ST/N000919/1. We gratefully acknowledge the use of matplotlib 2.0.0 \citep{matplotlib}, \textsc{naima} \citep{zabalza_naima:_2015} and \textsc{pluto} \citep{mignone_pluto:_2007}. This research made use of Astropy,\footnote{http://www.astropy.org} a community-developed core Python package for Astronomy \citep{astropy:2013, astropy:2018}.

\section*{References}

\end{document}